\documentclass[preprint,12pt,sort&compress]{elsarticle}

\usepackage{amsmath,amssymb,amsfonts}
\usepackage{graphicx}
\usepackage{booktabs}
\usepackage{multirow}
\usepackage{siunitx}
\usepackage{algorithm}
\usepackage{algpseudocode}
\usepackage{float}
\usepackage{url}
\usepackage{hyperref}
\usepackage{enumitem}
\usepackage{physics}
\usepackage{mathtools}
\usepackage{subcaption}
\usepackage{longtable}
\usepackage{lscape}
\usepackage{xurl}
\usepackage{microtype}

\journal{Applied Energy}

\begin{document}
	
	\begin{frontmatter}
		
		\title{An Integrated Techno-Economic Framework for Optimal Microgrid Design: An Australian Case Study}
		
		
		\author[inst1]{Mohamed Atef}
		\ead{Muhamed.Atef1408@gmail.com}
		
		\author[inst1]{Sanath Alahakoon}
		
		\author[inst2]{Umme Mumtahina}
		
		\author[inst2]{Peter Wolfs}
		
		\author[inst3]{Tamer Khatib}
		
		\author[inst4]{Moslem Uddin}
		\ead{moslem.uddin.bd@gmail.com}
		
		
		\address[inst1]{School of Engineering and Technology, Central Queensland University, Gladstone, QLD 4680, Australia}
		
		\address[inst2]{School of Engineering and Technology, Central Queensland University, Rockhampton, QLD 4680, Australia}
		
		\address[inst3]{Energy Engineering \& Environment Dept., An-Najah National University, Nablus, West Bank 97300, Palestine}
		
		\address[inst4]{School of Engineering \& Technology, The University of New South Wales, Canberra, ACT 2610, Australia}
		
		
		\begin{abstract}
			
			Reliable and affordable electricity supply remains a challenge for remote and regional communities, motivating the deployment of renewable-based microgrids supported by flexible storage and advanced planning methods. This paper proposes an integrated techno-economic framework for optimal microgrid design and robustness assessment, and applies it to a 1000-household residential community in Rockhampton, Queensland (Australia). The framework links time-series simulation, dispatch-based operation, and lifecycle costing to evaluate hybrid configurations comprising photovoltaic and wind generation, battery storage, diesel backup, grid exchange, and an optional hydrogen subsystem (electrolyzer--hydrogen storage--fuel cell).			
			Key indicators include net present cost (NPC), cost of energy (COE), renewable penetration, energy purchased/sold, and emissions-related outcomes. To avoid conclusions that depend on a single set of assumptions, the study performs systematic sensitivity analysis across financial, technical and policy drivers: discount rate, technology capital costs, fuel price, load uncertainty, renewable resource variability, carbon pricing/emissions cost, and grid outage duration, supplemented by a no-hydrogen attribution case.			
			The results demonstrate that several sensitivity dimensions induce nonlinear shifts in the optimal design, including breakpoints where capital-intensive renewable--storage expansion becomes economically preferable. The proposed framework enables transparent comparison of hydrogen-enabled and battery-centric solutions and provides planning guidance for resilient, low-emission community microgrids under Australian operating conditions.
			
		\end{abstract}
		
		
		
	\end{frontmatter}
	
	
	\section{Introduction}
Microgrids (MGs) have become increasingly prominent due to their ability to enhance energy reliability, particularly in isolated or distributed networks \cite{Guo2018, Li2019, Liu2019}. These systems incorporate diverse energy sources—such as solar, wind, and diesel generators—and can function either in connection with the main grid or independently \cite{gong2020secured,atef2020optimization,Li2023}. MGs contribute to improved power quality, uninterrupted supply to essential loads, and overall system performance \cite{hatziargyriou2007microgrids,gui2018passivity}. Nonetheless, the intermittent nature of renewable energy sources (RESs) and variable demand patterns introduce significant operational challenges \cite{ejuh2025impact,Che2017,RaiGoyal2024}.
Energy storage systems are essential for maintaining the balance between generation and consumption in MGs \cite{atef2019utilization,he2019small}. However, coordinating these systems with RESs and dynamic load profiles remains complex, prompting growing interest in microgrid energy management (MEM) strategies aimed at ensuring reliability and minimizing costs \cite{hao2019decentralized,atef2024data}.
Hydrogen-based technologies are gaining traction as a promising option for long-duration energy storage and carbon footprint reduction. For instance, \cite{elkhatib2023microchp} presents a solar-powered PEM fuel cell micro-CHP system in France that achieved complete electricity and heat coverage, with energy autonomy reaching 100\% depending on climatic conditions. Similarly, a study conducted in India \cite{das2024integration} utilized HOMER Pro to simulate hybrid RES-hydrogen configurations for residential and EV applications, showing notable improvements in cost efficiency and emission reductions.
In Calgary, a solar-hydrogen MG employing reversible solid oxide cells (rSOC) was analyzed in \cite{enaloui2023soec}. Serving 525 households, the system achieved monthly CO$_2$ reductions exceeding 250~kg per home, with feasibility enhanced by advancements in cell longevity and supportive policies. The work in \cite{islanded2023strategy} explored a net-zero islanded PV MG that integrated hydrogen components to fully meet energy demand and eliminate CO$_2$ emissions. Additionally, a Spanish case study \cite{spanish2023hybrid} combined PV, battery storage, and hydrogen systems to supply electricity and heat, enabling seasonal energy balancing and significant carbon mitigation.

\subsection{Background} 
Remote and sparsely populated regions of Australia still face persistent energy access and reliability constraints, driven by long distances, limited redundancy, and exposure to disruptive events. In fact, Australian rural/remote communities are often described as being more vulnerable to power outages, with reliability challenges exacerbated by hazards such as bushfires, which has strengthened the case for community-scale microgrids and localized energy autonomy \cite{uddin2023techno}. At the same time, the planning problem is not just “how to generate power,” but how to operate and coordinate resources under uncertainty (renewable variability, load fluctuations, component limits), which is why the literature consistently frames microgrids as systems that depend on effective energy management strategies to meet technical and economic objectives \cite{ali2023comprehensive,singh2023sustainable}.

Despite the momentum toward renewables, remote power supply remains largely diesel-dominated because diesel generation is dispatchable and operationally familiar in islanded settings; the same Australian context review notes that remote/off-grid generation “mainly relies on diesel generators,” which directly motivates decarbonization pathways \cite{uddin2023techno}. However, high renewable penetration introduces new operational and stability burdens, particularly as microgrids become more inverter-dominated, where stability and control coordination are recurrent technical challenges in the literature \cite{mohammadi2023stability,meraj2023energy}. As a result, the transition away from diesel is typically coupled with energy storage deployment and more advanced EMS: storage is repeatedly highlighted as critical for smoothing renewable intermittency and enabling higher renewable utilization, while EMS layers coordinate scheduling and real-time decisions \cite{ahmad2023ess,liu2022microgrid}. This complexity expands further in networked and multi-microgrid settings, where coordination and hierarchical control are often positioned as key enablers for resilience and efficient operation \cite{alvarez2023networked}.

In this broader transition, hydrogen-based systems are increasingly proposed as a pathway to reduce diesel dependency and support deep decarbonization in islanded microgrids. Hydrogen microgrids typically integrate renewables with electrolyzers, hydrogen storage, and fuel cells, and the hydrogen-focused review literature emphasizes that robust and cost-effective operation depends strongly on the EMS, because hydrogen production, storage, and reconversion must be coordinated with variable renewables and load demands \cite{van2023review}. At the system level, a common argument is that batteries and hydrogen are complementary: batteries handle short-term peaks and fast dynamics, while hydrogen offers long-duration storage without self-discharge—so the “best of both” only happens when the EMS explicitly manages their synergy \cite{naseri2024energy}. Recent techno-economic studies similarly frame hydrogen as attractive for medium-to-large off-grid applications where battery cost/constraints become limiting, with seasonal hydrogen storage improving feasibility in off-grid supply\cite{enaloui2025techno}. Given the uncertainty in renewable resources, demand growth, and cost assumptions, the literature also stresses the importance of forecasting and scenario-based evaluation in multi-microgrid energy management \cite{raj2025review}. Consequently, techno-economic planning studies frequently rely on tools that support scenario and sensitivity analysis; for example, the microgrid planning review explicitly notes that the HOMER tool is widely used for hybrid-system simulation and includes sensitivity analysis capabilities—directly aligning with this paper’s focus on an Australian case study supported by HOMER-based sensitivity assessment \cite{AghaKassab2024}.

According to \cite{van2023review}, hydrogen-based microgrids are increasingly framed as renewable-driven systems where surplus electricity is converted via electrolysers, stored as hydrogen, and later reconverted via fuel cells—and the energy management system (EMS) is therefore a critical enabler of techno-economic optimal operation.

\subsection{Motivation and gap} 

The motivation for this work is driven by the reality that remote and weakly supported power systems must satisfy reliability, affordability, and resilience requirements while operating under high uncertainty in renewable output and demand. In the Australian context, community/remote microgrids are increasingly discussed as a practical pathway to improve local supply security and resilience, particularly where conventional grid expansion is uneconomic or operationally fragile\cite{uddin2023techno}. Beyond single microgrids, the need becomes more pressing in multi-microgrid / networked microgrid settings: fluctuating weather patterns and variable consumption behaviors make it difficult to maintain dependable supply, motivating tighter integration between EMS and forecasting for scheduling and unit commitment\cite{raj2025review}. From a system viewpoint, networked microgrids are positioned to improve stability and resilience via coordination and energy sharing, but this introduces additional layers of uncertainty, privacy/security concerns, and cooperative operation challenges that are still actively being addressed \cite{rodriguez2024energy}. 

At the same time, many remote/off-grid deployments remain diesel-dominated, largely because diesel generators provide dispatchable capacity and operational familiarity. However, diesel reliance brings clear economic and environmental burdens—particularly fuel logistics and emissions—reinforcing the push toward renewable-dominant architectures supported by advanced control and storage\cite{uddin2023microgrids}. As renewable penetration increases, microgrids become more inverter-dominated, which is repeatedly linked to new stability and protection challenges and a stronger need for robust EMS and coordination\cite{alvarez2023networked}. Recent stability-focused work further emphasizes that modern microgrids face rising complexity (RES integration, EV loads, cyber-physical exposure), and that AI-driven EMS is promising—yet real-world deployment and validation remain limited in many cases\cite{mohammadi2023stability}. 

Despite substantial progress, a persistent research gap is the disconnect between techno-economic design/planning and operationally realistic EMS under uncertainty. Planning studies frequently rely on well-known tools and standardized workflows, but reviews still highlight limitations related to uncertainty representation, modelling depth, and translating design outputs into robust operating strategies—especially when the system must be resilient to variability in solar/wind resources, load growth, and price assumptions\cite{AghaKassab2024}. In parallel, techno-economic case studies repeatedly show that sensitivity analysis is not optional: it is used to test whether an “optimal” design remains viable when key drivers shift (resource availability, demand, prices, technology costs). For example, Australian multi-energy community microgrid studies explicitly examine uncertainty drivers (e.g., diesel prices, demand growth, discount rate) through sensitivity analysis, reinforcing that robustness is central to decision credibility\cite{uddin2023techno}. Similar sensitivity-driven evaluation is highlighted as important in multi-microgrid contexts where forecasting and uncertainty directly affect scheduling and reliability outcomes\cite{raj2025review}. 

The gap becomes sharper when the design expands to include hydrogen as a long-duration flexibility pathway. Hydrogen-microgrid reviews emphasize that EMS selection must account for objectives (technical/economic/ environmental), constraints, and configuration, and they explicitly identify under-addressed needs such as forecasting integration, environmental objectives, and accurate modelling with real-life constraints (e.g., grid congestion assumptions) \cite{van2023review,atef2025techno}. The same review highlights that hydrogen is commonly paired with short-term storage (e.g., batteries), implying that practical frameworks should co-optimize both short- and long-duration storage roles rather than treating hydrogen as a bolt-on component\cite{van2023review,atef2025real}. Finally, broader trend analyses argue that limited real-world validation, the need for coordinated control in interconnected systems, and supportive policy/interconnection mechanisms remain barriers to scalable deployment—particularly relevant for remote, high-stakes energy systems\cite{gonzalez2025exploring}.

\subsection{Contributions}
Building on the identified gaps between techno-economic planning and operationally realistic energy management under uncertainty, \cite{AghaKassab2024, abbasi2023recent}, and responding to the need for stronger forecasting/uncertainty integration \cite{raj2025review}, and networked/coordination-aware EMS \cite{rodriguez2024energy}, this paper makes the following contributions:
\begin{itemize}
    \item Integrated techno-economic framework (design → operation → robustness):
We propose an end-to-end framework that links (i) Australian case-study data preparation, (ii) optimal system sizing and dispatch assessment, and (iii) robustness evaluation via structured sensitivity analysis—addressing calls for more transparent, uncertainty-aware microgrid planning workflows \cite{AghaKassab2024,gonzalez2025exploring}.
    \item Australian decision-relevant case study with remote-energy context:
We ground the framework in an Australian case study to reflect the practical constraints of remote and regional supply, where microgrids are increasingly considered to improve reliability and autonomy \cite{uddin2023techno, uddin2023microgrids}.
    \item Comparative assessment of multiple microgrid architectures (including hydrogen):
We evaluate and compare feasible architecture options (renewables + storage + backup) and explicitly include hydrogen-based pathways as long-duration flexibility, consistent with the growing emphasis on hydrogen-enabled microgrids and their EMS requirements \cite{van2023review,modu2023systematic}.
    \item HOMER-driven sensitivity analysis designed for planning robustness (not just “best-case” sizing):
Beyond reporting a single optimal configuration, we use HOMER sensitivity analysis to quantify how key uncertainties (e.g., technology costs, fuel prices, demand/tariffs) alter the preferred design and the economic/emissions trade-offs—aligning with the literature’s emphasis that robustness and sensitivity testing are essential in credible microgrid planning \cite{AghaKassab2024,sackey2023techno}.
    \item Storage-centric operational insights (battery vs hydrogen roles):
We provide a structured interpretation of how storage technologies support renewable penetration—batteries for fast dynamics and short-term balancing, hydrogen for extended-duration adequacy—reflecting the broader view that hybrid storage portfolios require explicit coordination for performance and cost-effectiveness \cite{ahmad2023ess,naseri2024energy}.
    \item Actionable design guidance derived from cross-metric trade-offs:
We translate outputs into decision ready recommendations by jointly reporting economic metrics (e.g., NPC/COE), operational feasibility indicators, and emissions outcomes supporting the “framework-to-decision” expectation increasingly highlighted in EMS and microgrid planning literature \cite{abbasi2023recent,ahmed2023techno}.
    \item Future-proofing alignment with emerging EMS directions:
The proposed framework is structured to be extendable toward forecasting-aware scheduling, advanced optimization/control, and resilience-aware operation—consistent with trends identified across recent EMS and stability literature \cite{raj2025review,mohammadi2023stability}.

\end{itemize}


\subsection{Paper Structure}
The remainder of this paper is structured as follows. Section~2 presents the overall system description and the proposed techno-economic methodology, including the case-study boundary, component modelling, optimization/dispatch workflow, evaluation metrics, and the sensitivity analysis design. Section~3 reports the results and discussion, beginning with the optimal configuration and cost analysis, followed by environmental and operational performance indicators, and then a comprehensive sensitivity assessment covering discount rate, technology capital cost, fuel price, load demand uncertainty, renewable resource variability, carbon pricing and emissions cost, grid outage duration, and a no-hydrogen attribution case. Section~4 discusses the broader implications of the findings for resilient and low-emission community microgrids and highlights practical policy considerations. Finally, Section~5 concludes the paper and outlines key takeaways and directions for future work.

\section{System Description and Methodology}\label{sec:method}

This paper upgrades the prior conference methodology into a journal-grade, decision-oriented framework for optimal microgrid (MG) design under uncertainty. The proposed workflow follows best practices recommended in recent microgrid planning and energy-management literature, including: (i) explicit definition of the system boundary, (ii) chronological simulation to capture renewable variability and storage dynamics, (iii) lifecycle techno-economic and emissions evaluation, and (iv) structured sensitivity analysis to quantify robustness and avoid over-reliance on a single point-optimal solution \cite{atef2026energy,wang2024emsreview,fernandez2025distribution}.

\subsection{System description and case study boundary (1000-household model)}\label{subsec:case}
The study considers a residential community microgrid in Rockhampton, Queensland (Australia), extending the baseline conference case to a larger planning scale by modeling an aggregated community of $N_{\mathrm{hh}}=1000$ households \cite{atef2025techno,AghaKassab2024}. The MG boundary includes generation units, energy storage systems (ESS), power conversion devices, optional hydrogen conversion/storage, and grid exchange (when available), consistent with contemporary microgrid planning formulations \cite{uddin2023microgrids, rodriguez2024energy}.

If $P_{\mathrm{hh}}(t)$ denotes the representative per-household electrical demand (kW) at time $t$, then the aggregated community load is:
\begin{equation}
P_{L}(t)=N_{\mathrm{hh}}\,P_{\mathrm{hh}}(t), \qquad N_{\mathrm{hh}}=1000 .
\label{eq:load_aggregate}
\end{equation}
where $P_{L}(t)$ is the total community electrical demand (kW), $N_{\mathrm{hh}}$ is the number of households (dimensionless), and $t$ denotes the simulation time index (e.g., hourly).

If an aggregated base profile $P_{\mathrm{base}}(t)$ exists for $N_{\mathrm{base}}$ households, scaling to 1000 households is expressed as:
\begin{equation}
P_{L}(t)=\left(\frac{1000}{N_{\mathrm{base}}}\right)P_{\mathrm{base}}(t),
\label{eq:load_scaling}
\end{equation}
where $P_{\mathrm{base}}(t)$ is the base aggregated demand (kW) and $N_{\mathrm{base}}$ is the base household count (dimensionless). Such scaling is commonly used in community planning when the objective is system-level sizing and cost robustness \cite{AghaKassab2024, zheng2024systematic}.

\subsection{Input datasets: load, tariff, and renewable resources}\label{subsec:inputs}
The simulation employs (i) an aggregated residential load profile (Section~\ref{subsec:case}), (ii) a time-of-use electricity tariff structure consistent with the baseline conference workflow to maintain comparability across scenarios \cite{atef2025techno,AghaKassab2024}, and (iii) location-specific renewable resource time series (solar irradiance, wind speed, ambient temperature). Chronological simulation is adopted because it is widely recognized as necessary to capture the impacts of intermittency, storage cycling, and dispatch feasibility in hybrid microgrids \cite{AghaKassab2024, ferahtia2024recent, wang2024emsreview}.

\subsection{Candidate configurations and energy assets}\label{subsec:assets}
Two coupled pathways are considered:
\begin{itemize}
	\item \textbf{(A) Electric pathway:} photovoltaic (PV), wind turbine(s) (WT), battery energy storage system (BESS), and inverter/converter, supported by diesel generator(s) (DG) and/or grid exchange depending on scenario constraints \cite{AghaKassab2024, ali2023applications, uddin2023microgrids}.
	
	\item \textbf{(B) Hydrogen pathway (optional):} electrolyzer, hydrogen storage tank, and fuel cell (FC), enabling long-duration energy shifting and improved autonomy under renewable deficits \cite{ van2023review, modu2023systematic, sadeghian2025energy}.
\end{itemize}

This hybrid storage framing is consistent with recent EMS literature where batteries provide short-term balancing while hydrogen provides longer-duration flexibility, provided EMS coordination is appropriately designed \cite{ van2023review, sadeghian2025energy}.

\subsection{Component modelling}\label{subsec:models}
The MG is modeled using standard energy-balance relations suitable for planning-grade techno-economic analysis \cite{AghaKassab2024, fernandez2025distribution, atef2026review}.

\subsubsection{PV generation model}
PV output is represented as:
\begin{equation}
P_{\mathrm{PV}}(t)=P_{\mathrm{PV,r}}\,f_{\mathrm{der}}\,
\frac{G(t)}{G_{\mathrm{STC}}}\left[1+\alpha_{p}\left(T_{c}(t)-T_{\mathrm{STC}}\right)\right],
\label{eq:pv}
\end{equation}
where $P_{\mathrm{PV}}(t)$ is PV electrical output (kW), $P_{\mathrm{PV,r}}$ is rated PV capacity (kW), $f_{\mathrm{der}}$ is PV derating factor (dimensionless), $G(t)$ is plane-of-array irradiance (W/m$^2$), $G_{\mathrm{STC}}$ is irradiance at standard test conditions (W/m$^2$), $\alpha_{p}$ is PV power temperature coefficient (1/$^\circ$C), $T_{c}(t)$ is PV cell temperature ($^\circ$C), and $T_{\mathrm{STC}}$ is cell temperature at STC ($^\circ$C). This formulation is consistent with the baseline conference modelling approach and planning literature \cite{atef2025techno, AghaKassab2024}.

A common approximation for PV cell temperature is:
\begin{equation}
T_{c}(t)=T_{a}(t)+\left(\frac{\mathrm{NOCT}-20}{800}\right)G(t),
\label{eq:celltemp}
\end{equation}
where $T_{a}(t)$ is ambient temperature ($^\circ$C) and $\mathrm{NOCT}$ is nominal operating cell temperature ($^\circ$C) \cite{atef2026energy, AghaKassab2024}.

\subsubsection{Wind turbine model}
Wind output can be expressed using an aerodynamic form (with practical implementation via manufacturer power curves):
\begin{equation}
P_{\mathrm{WT}}(t)=\frac{1}{2}\rho\,A\,C_{p}\,v(t)^{3},
\label{eq:wind}
\end{equation}
where $P_{\mathrm{WT}}(t)$ is WT electrical output (kW), $\rho$ is air density (kg/m$^3$), $A$ is rotor swept area (m$^2$), $C_{p}$ is power coefficient (dimensionless), and $v(t)$ is hub-height wind speed (m/s). This model is aligned with common WT representations in planning and EMS surveys \cite{atef2025techno,AghaKassab2024,ferahtia2024recent}.

\subsubsection{Battery energy storage model}
The BESS state of charge (SOC) is updated as:
\begin{align}
SOC(t+\Delta t) &= SOC(t)+\frac{\eta_{\mathrm{ch}}\,P_{\mathrm{ch}}(t)\,\Delta t}{E_{\mathrm{bat}}}, \quad \text{(charging)} \label{eq:soc_charge}\\
SOC(t+\Delta t) &= SOC(t)-\frac{P_{\mathrm{dis}}(t)\,\Delta t}{\eta_{\mathrm{dis}}\,E_{\mathrm{bat}}}, \quad \text{(discharging)} \label{eq:soc_dis}
\end{align}
subject to
\begin{equation}
SOC_{\min}\le SOC(t)\le SOC_{\max},
\label{eq:soc_bounds}
\end{equation}
where $SOC(t)$ is battery state of charge (dimensionless or \%), $\Delta t$ is time-step duration (h), $\eta_{\mathrm{ch}}$ is charging efficiency (dimensionless), $\eta_{\mathrm{dis}}$ is discharging efficiency (dimensionless), $P_{\mathrm{ch}}(t)$ is battery charging power (kW), $P_{\mathrm{dis}}(t)$ is battery discharging power (kW), $E_{\mathrm{bat}}$ is battery energy capacity (kWh), and $SOC_{\min}$ and $SOC_{\max}$ are the minimum and maximum permissible SOC limits (dimensionless or \%). Such SOC-constrained formulations are standard in microgrid EMS/control studies \cite{atef2025techno, chen2025dc, eyimaya2024review,AghaKassab2024}.

\subsubsection{Diesel generator fuel model}
A commonly used linear fuel model is:
\begin{equation}
F(t)=a\,P_{\mathrm{DG}}(t)+b\,P_{\mathrm{DG,r}},
\label{eq:diesel_fuel}
\end{equation}
where $F(t)$ is fuel consumption rate (e.g., L/h), $a$ is fuel-slope coefficient (L/kWh), $P_{\mathrm{DG}}(t)$ is DG output power (kW), $b$ is fuel-intercept coefficient (L/h per kW-rated or equivalent representation), and $P_{\mathrm{DG,r}}$ is DG rated capacity (kW). This representation is frequently adopted in techno-economic planning, particularly in studies assessing fuel-price sensitivity \cite{AghaKassab2024, sackey2023techno,ahmed2023techno}.

\subsubsection{Hydrogen subsystem model (electrolyzer--storage--fuel cell)}
Electrolyzer hydrogen production (energy basis) is:
\begin{equation}
E_{H_{2}}^{\mathrm{prod}}(t)=\eta_{\mathrm{el}}\,P_{\mathrm{el}}(t)\,\Delta t,
\label{eq:h2_prod}
\end{equation}
Hydrogen storage balance is:
\begin{equation}
E_{H_{2}}(t+\Delta t)=E_{H_{2}}(t)+E_{H_{2}}^{\mathrm{prod}}(t)-E_{H_{2}}^{\mathrm{cons}}(t),
\label{eq:h2_balance}
\end{equation}
Fuel cell output is:
\begin{equation}
P_{\mathrm{FC}}(t)=\eta_{\mathrm{fc}}\frac{E_{H_{2}}^{\mathrm{cons}}(t)}{\Delta t},
\label{eq:fc}
\end{equation}
subject to:
\begin{equation}
E_{H_{2},\min}\le E_{H_{2}}(t)\le E_{H_{2},\max},
\label{eq:h2_bounds}
\end{equation}
where $\eta_{\mathrm{el}}$ is electrolyzer efficiency (dimensionless), $P_{\mathrm{el}}(t)$ is electrolyzer electrical input power (kW), $E_{H_{2}}^{\mathrm{prod}}(t)$ is hydrogen energy produced in the time step (kWh$_{H_2}$), $E_{H_{2}}(t)$ is stored hydrogen energy (kWh$_{H_2}$), $E_{H_{2}}^{\mathrm{cons}}(t)$ is hydrogen energy consumed by the fuel cell (kWh$_{H_2}$), $\eta_{\mathrm{fc}}$ is fuel cell electrical efficiency (dimensionless), $P_{\mathrm{FC}}(t)$ is fuel cell electrical output (kW), and $E_{H_{2},\min}$ and $E_{H_{2},\max}$ are minimum and maximum hydrogen storage bounds (kWh$_{H_2}$). These relations capture the conversion losses and capacity constraints emphasized in hydrogen-microgrid EMS literature \cite{van2023review, modu2023systematic, sadeghian2024energy}.

\subsubsection{System power balance}
At each time step, the microgrid power balance is:
\begin{align}
&P_{\mathrm{PV}}(t)+P_{\mathrm{WT}}(t)+P_{\mathrm{DG}}(t)+P_{\mathrm{FC}}(t)+P_{\mathrm{grid,in}}(t)+P_{\mathrm{dis}}(t) \nonumber\\
&=P_{L}(t)+P_{\mathrm{ch}}(t)+P_{\mathrm{el}}(t)+P_{\mathrm{grid,out}}(t),
\label{eq:power_balance}
\end{align}
where $P_{\mathrm{grid,in}}(t)$ is grid import power (kW), $P_{\mathrm{grid,out}}(t)$ is grid export power (kW), and all other terms are defined in Eqs.~\eqref{eq:pv}--\eqref{eq:fc}. Such balance constraints are standard in EMS and active distribution network formulations \cite{ fernandez2025distribution, eyimaya2024review}.

\subsection{Optimization and dispatch methodology}\label{subsec:dispatch}
The sizing and operational assessment follow a simulation--optimization workflow in which candidate configurations are simulated chronologically under dispatch and ranked by lifecycle techno-economic performance \cite{atef2025techno,AghaKassab2024}. Dispatch prioritizes renewable utilization, then allocates surplus to storage (battery and/or hydrogen), while deficits are covered through storage discharge, grid import (if available), and dispatchable backup. The need for coordinated EMS under renewable variability and inverter-dominated operation is emphasized in recent EMS and stability surveys \cite{ abbasi2023recent, safder2023enhancing, rodriguez2024energy, han2024hierarchical}.

\subsection{Evaluation metrics}\label{subsec:metrics}
The primary economic objective is Net Present Cost (NPC), supported by Cost of Energy (COE/LCOE) and emissions indicators, consistent with contemporary microgrid planning literature \cite{atef2026energy, uddin2023microgrids}.

The capital recovery factor (CRF) is:
\begin{equation}
CRF(i,N)=\frac{i(1+i)^{N}}{(1+i)^{N}-1},
\label{eq:crf}
\end{equation}
where $i$ is the annual real discount rate (dimensionless) and $N$ is the project lifetime (years).

NPC can be computed from the total annualized cost $C_{\mathrm{ann}}$ as:
\begin{equation}
NPC=\frac{C_{\mathrm{ann}}}{CRF(i,N)},
\label{eq:npc}
\end{equation}
where $C_{\mathrm{ann}}$ is the annualized total cost (currency/year), including annualized capital cost, replacement cost, fixed and variable O\&M, fuel cost, and net grid purchase cost (as applicable).

COE is computed as:
\begin{equation}
COE=\frac{C_{\mathrm{ann}}}{E_{\mathrm{served}}},
\label{eq:coe}
\end{equation}
where $E_{\mathrm{served}}$ is the annual electrical energy served to the load (kWh/year).

Carbon pricing impact (if applied) can be represented as:
\begin{equation}
C_{\mathrm{CO_2}}=p_{\mathrm{CO_2}}\,E_{\mathrm{CO_2}},
\label{eq:carbon_cost}
\end{equation}
where $C_{\mathrm{CO_2}}$ is the annual carbon cost (currency/year), $p_{\mathrm{CO_2}}$ is carbon price (currency per tCO$_2$), and $E_{\mathrm{CO_2}}$ is annual CO$_2$ emissions (tCO$_2$/year). Carbon/emissions cost signals are increasingly recognized as planning drivers, especially in multi-energy community microgrids \cite{uddin2023techno}.

\subsection{Sensitivity analysis design}\label{subsec:sensitivity}
A structured sensitivity analysis is performed to quantify robustness and identify dominant uncertainty drivers. The need for sensitivity testing is repeatedly emphasized in techno-economic microgrid planning studies and case analyses because parameter uncertainty can shift the optimal architecture and economics \cite{ sackey2023techno, ahmed2023techno, AghaKassab2024}. Sensitivity dimensions include (but are not limited to) technology capital costs, fuel price, load scaling, renewable resource variability, discount rate, carbon pricing, and grid outage conditions; similar uncertainty categories and forecasting-aware planning motivations are discussed in multi-microgrid EMS surveys \cite{ raj2025review, xu2024overview}.


\section{Results and Discussion}
\subsection{Optimal Configuration and Cost Analysis}
The baseline (reference) case is taken from the full-parameter model with fuel price multiplier equal to 1.0 (consistent across the sensitivity exports). Table~\ref{tab:base_econ} reports the key techno-economic outputs, while Table~\ref{tab:base_sizes} lists the corresponding optimal installed capacities/sizes (as reported by the model).

\begin{table}[!t]
\centering
\caption{Baseline techno-economic performance (reference case).}
\label{tab:base_econ}
\begin{tabular}{rrrr}
\toprule
NPC (A\$M) & COE (A\$/kWh) & Total CAPEX (A\$M) & OpEx (A\$k/yr) \\
\midrule
11.7922 & 0.0596 & 1.4408 & 330.3888 \\
\bottomrule
\end{tabular}%
\end{table}

\begin{table}[!t]
\centering
\caption{Baseline optimal system sizing (reference case).}
\label{tab:base_sizes}
\resizebox{\linewidth}{!}{%
\begin{tabular}{rrrrrrrr}
\toprule
 PV  &  Wind  &  Battery  &  Diesel  &  Electrolyzer  &  Fuel cell  &  H2 tank  &  Inverter  \\
 (kW)&(kW)&(kW)&(kW)&(kW)&(kW)&(kg)&(kW)\\
\midrule
 2055.594 & 138.000 & 5300.000 & 180.000 & 208.000 & 135.000 & 1370.000 & 614.000 \\
\bottomrule
\end{tabular}%
}
\end{table}

\subsection{Environmental and Operational Performance}
Table~\ref{tab:base_ops} summarizes the operational energy exchange and the overall renewable penetration for the baseline configuration. Negative net CO$_2$ values indicate that exported electricity offsets a larger amount of grid-related emissions than the residual emissions attributed to imports and backup operation within the accounting approach used by the simulation export.

\begin{table}[!t]
\centering
\caption{Baseline operational and emissions indicators (reference case).}
\label{tab:base_ops}
\begin{tabular}{rrrr}
\toprule
  RP (\%) &  EP (MWh/yr) &  ES (MWh/yr) &  CO2 (t/yr) \\
\midrule
 109.721 & 13855.180 & 2792.024 & -1651.873 \\
\bottomrule
\end{tabular}%
\end{table}

\subsection{Sensitivity Analysis}




\subsubsection{Discount Rate Sensitivity}
\begin{figure*}[!tp]
\centering

\begin{subfigure}{0.32\textwidth}
    \centering
    \includegraphics[width=\linewidth]{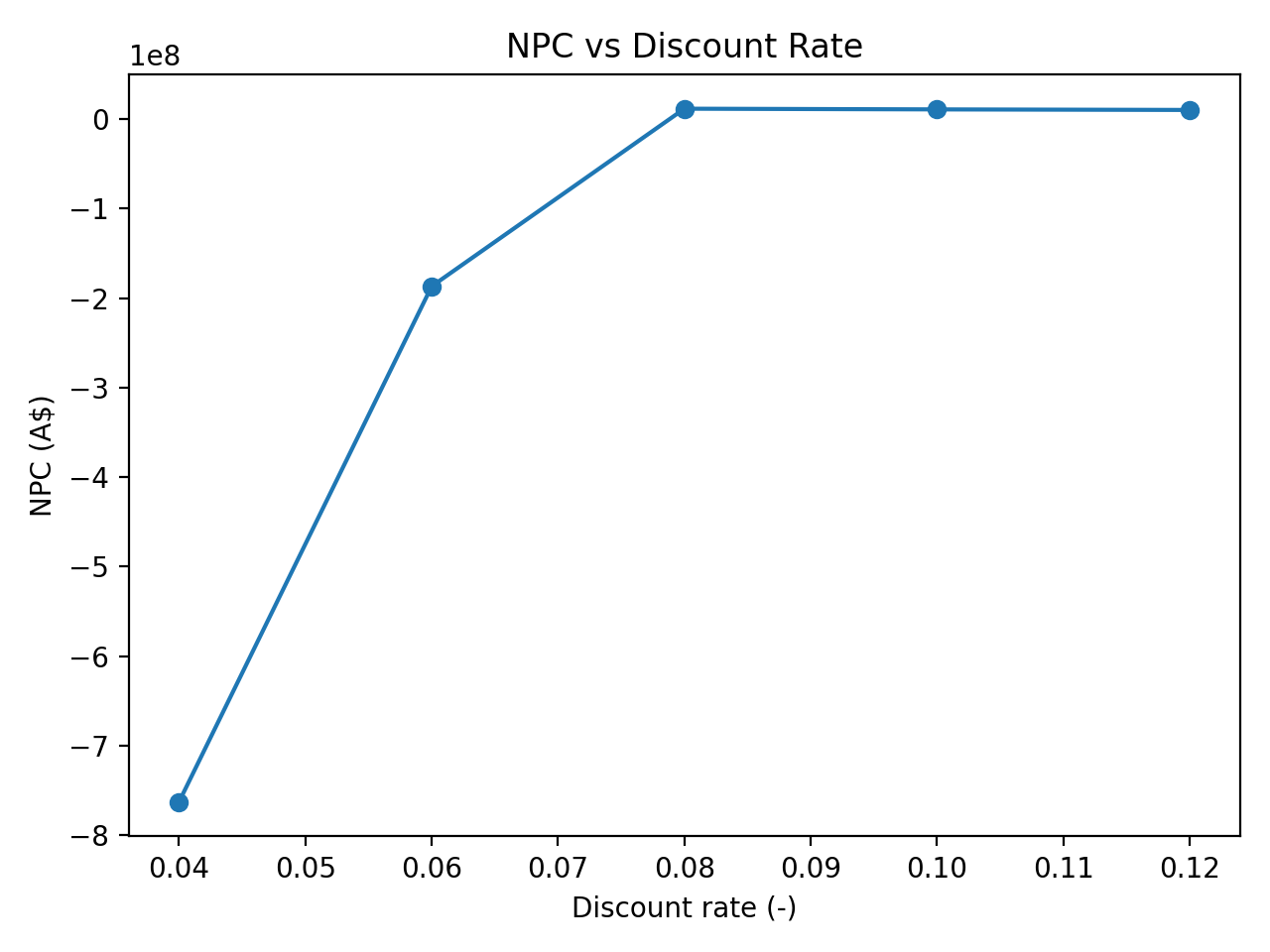}
    \caption{
    }
    \label{fig:dr_npc}
\end{subfigure}
\hfill
\begin{subfigure}{0.32\textwidth}
    \centering
    \includegraphics[width=\linewidth]{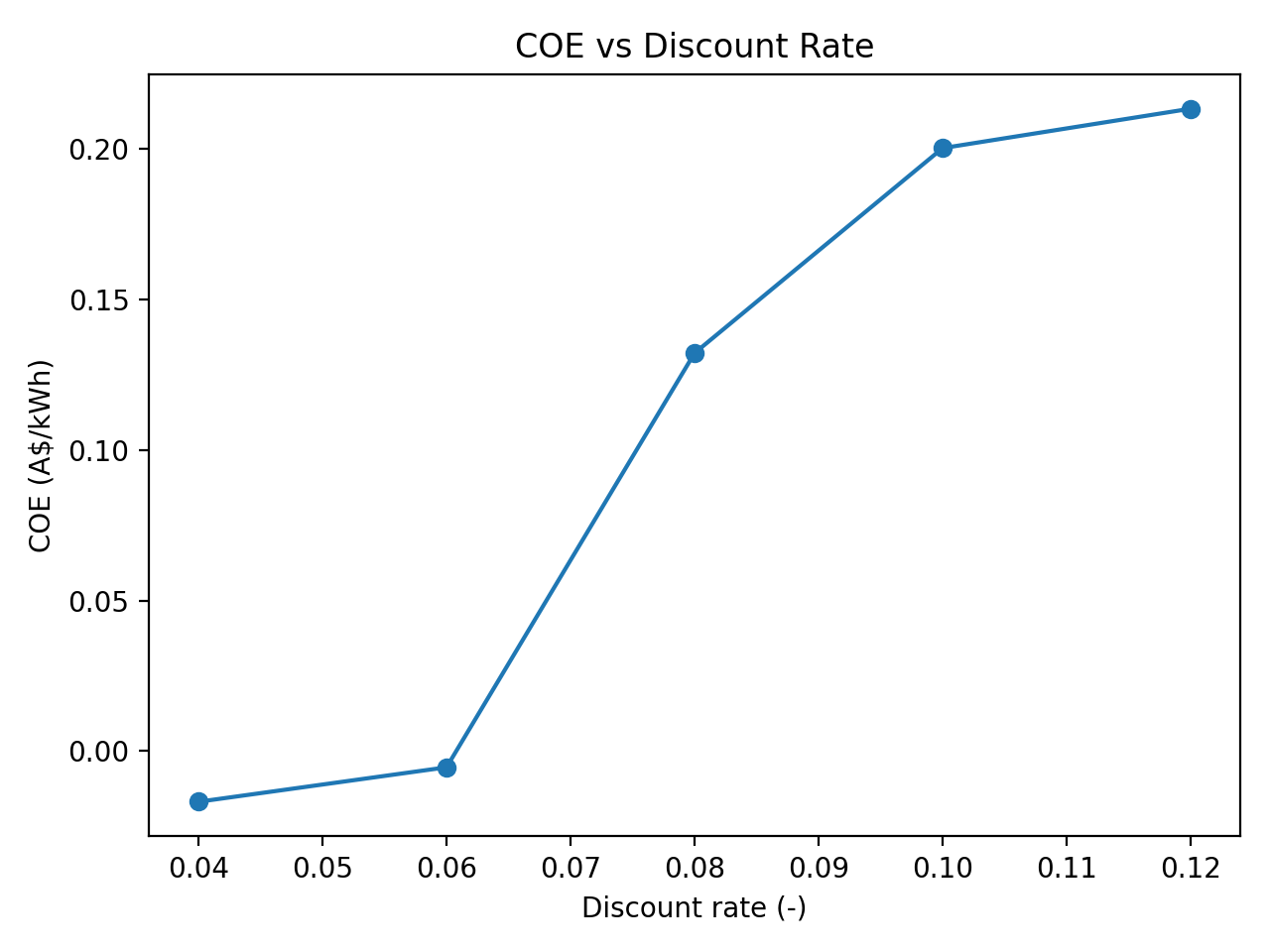}
    \caption{
    }
    \label{fig:dr_coe}
\end{subfigure}
\hfill
\begin{subfigure}{0.32\textwidth}
    \centering
    \includegraphics[width=\linewidth]{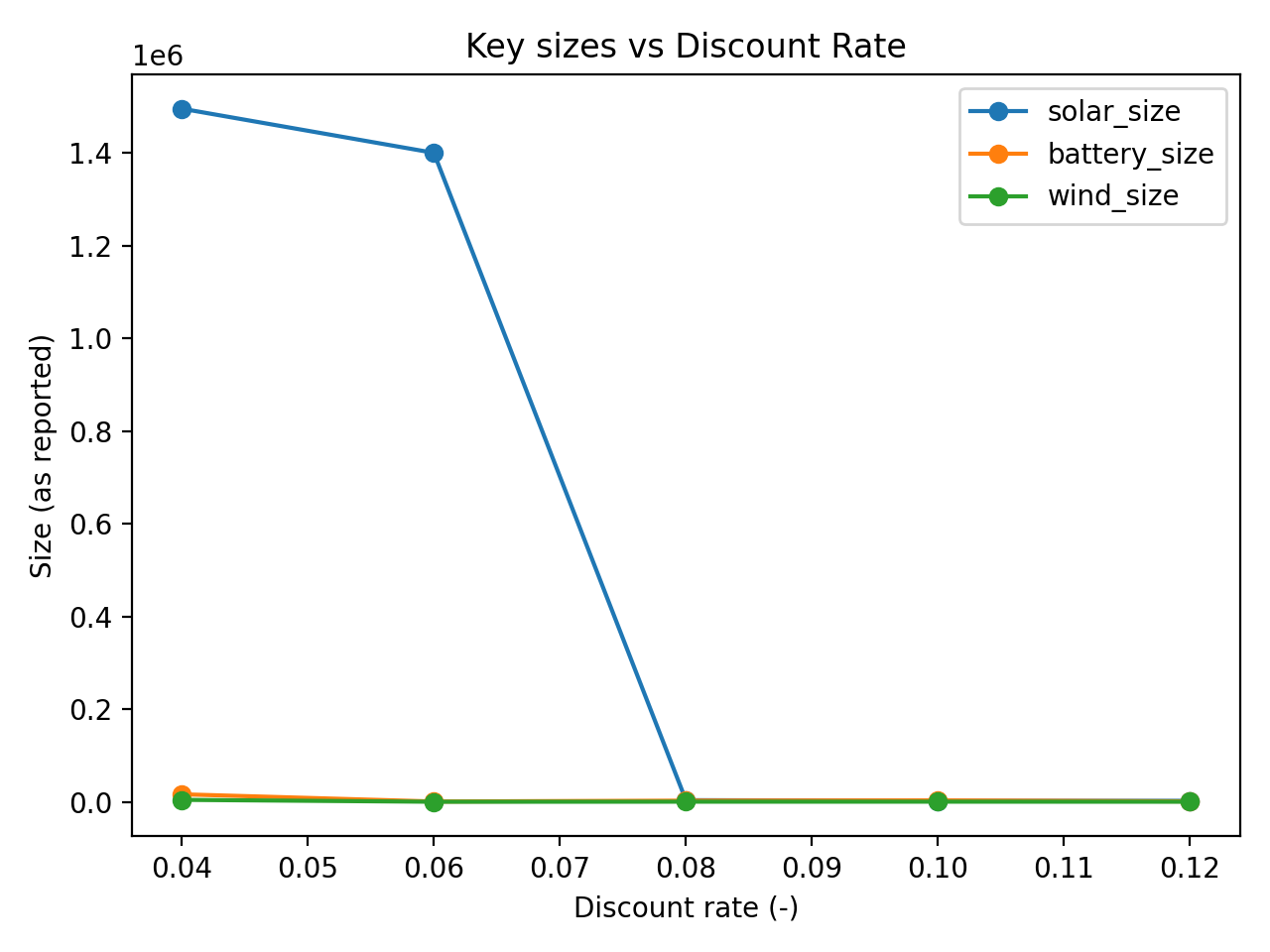}
    \caption{
    }
    \label{fig:dr_sizes}
\end{subfigure}

\caption{
Effect of discount rate on techno-economic performance and system sizing:
(a) NPC as a function of discount rate.
(b) COE under varying discount rates.
(c) Optimal component capacities obtained for different discount-rate assumptions.}
\label{fig:discount_rate_results}

\end{figure*}

Tables~\ref{tab:dr_econ}--\ref{tab:dr_sizes} and Fig.~\ref{fig:dr_npc}--\ref{fig:dr_sizes} show how the discount rate changes the lifecycle economics and the resulting sizing. The results exhibit a clear breakpoint: at very low discount rates (4--6\%), the optimizer selects extremely large renewable capacities and the NPC becomes negative, indicating an export-revenue/arbitrage-dominated solution. For decision realism, the 8--12\% range provides more practical designs, where higher discount rates reduce capital-intensive renewable deployment (lower renewable penetration) and shift the balance toward lower NPC but higher COE due to reduced export benefits.

\begin{table*}[!t]
\centering
\caption{Discount rate sensitivity: economic outputs.}
\label{tab:dr_econ}
\resizebox{\columnwidth}{!}{%
\begin{tabular}{rrrrrrr}
\toprule
 Discount rate &  NPC (A\$M) &  COE (A\$/kWh) &  CAPEX (A\$M) &  OpEx (A\$k/yr) &  Ren. pen. (\%) &  CO2 (t/yr) \\
\midrule
        0.0400 &  -66.8951 &      -0.3177 &       12.5530 &     -8432.6841 &       205.7568 & -13632.7830 \\
        0.0600 & -252.0367 &      -1.2320 &        8.3196 &    -27432.7669 &       234.1461 &  -8702.1960 \\
        0.0800 &   47.3014 &       0.1454 &        1.8506 &      1232.3534 &       106.7186 &  -2268.0620 \\
        0.1000 &   20.6427 &       0.0852 &        1.2349 &       611.3358 &       110.8513 &  -1367.4910 \\
        0.1200 &   15.2405 &       0.0718 &        1.0347 &       434.7717 &        91.5576 &   -862.3550 \\
\bottomrule
\end{tabular}
}
\end{table*}

\begin{table*}[!t]
\centering
\caption{Discount rate sensitivity: optimal sizes.}
\label{tab:dr_sizes}
\resizebox{\columnwidth}{!}{%
\begin{tabular}{rrrrrrrr}
\toprule
 Discount rate &      PV &   Wind &  Battery &  Diesel &  Electrolyzer &  Fuel cell &  H2 tank \\
\midrule
        0.0400 & 24985.800 &  26.000 & 5300.000 & 180.000 &     8065.000 &   135.000 & 1400.000 \\
        0.0600 & 15286.000 &  32.000 & 6200.000 & 180.000 &     2520.000 &   135.000 & 2200.000 \\
        0.0800 &  2491.594 & 138.000 & 5300.000 & 180.000 &      250.000 &   135.000 & 1370.000 \\
        0.1000 &  1797.035 & 110.000 & 5300.000 & 180.000 &      156.000 &   135.000 & 1520.000 \\
        0.1200 &  1223.000 & 100.000 & 2600.000 & 180.000 &      232.000 &   135.000 & 1450.000 \\
\bottomrule
\end{tabular}
}
\end{table*}




\subsubsection{Technology Capital Cost Sensitivity}
\begin{figure*}[!t]
	\centering
	
	\begin{subfigure}{0.48\textwidth}
		\centering
		\includegraphics[width=\linewidth]{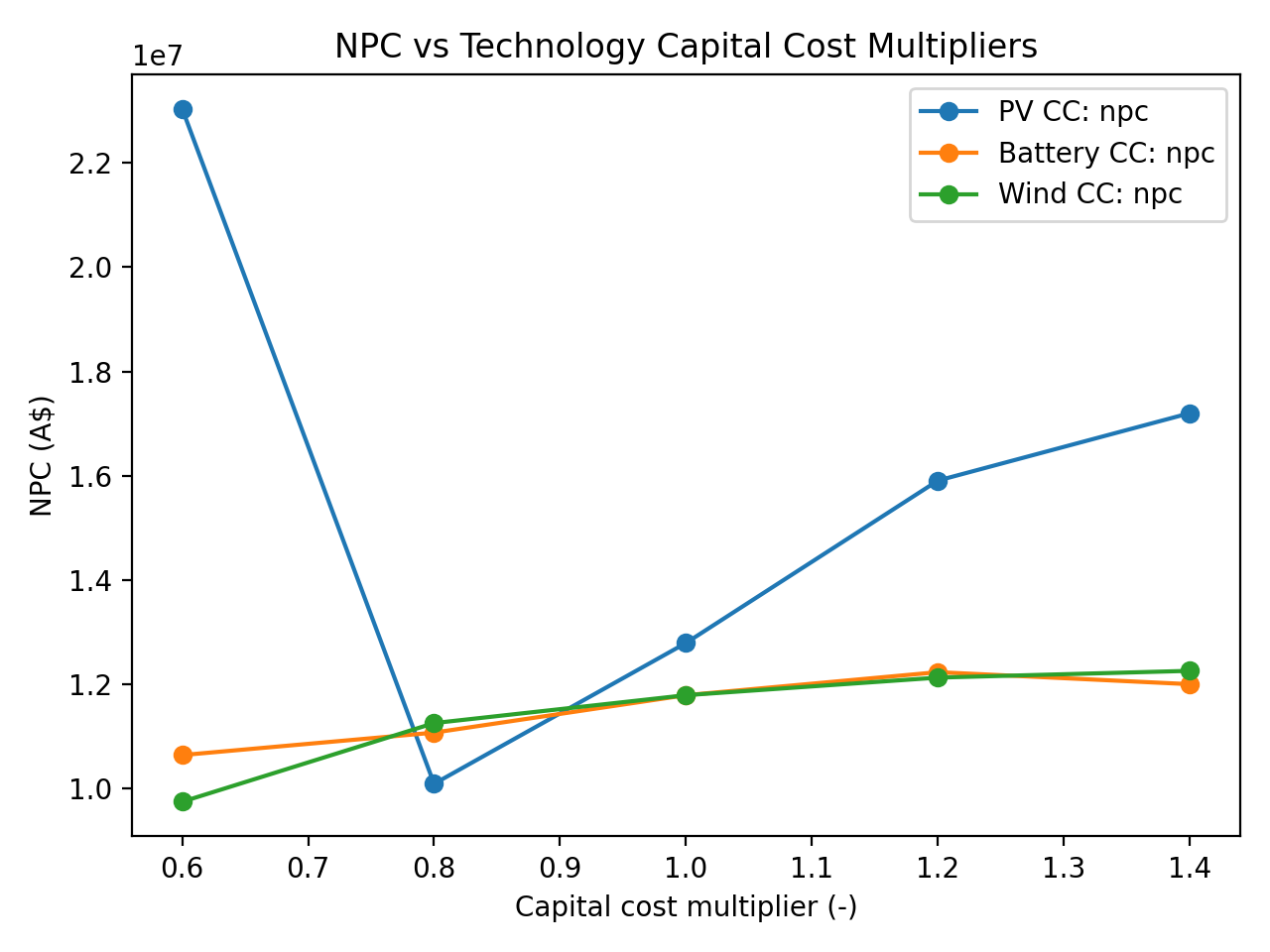}
		\caption{}
		\label{fig:cc_npc}
	\end{subfigure}
	\hfill
	\begin{subfigure}{0.48\textwidth}
		\centering
		\includegraphics[width=\linewidth]{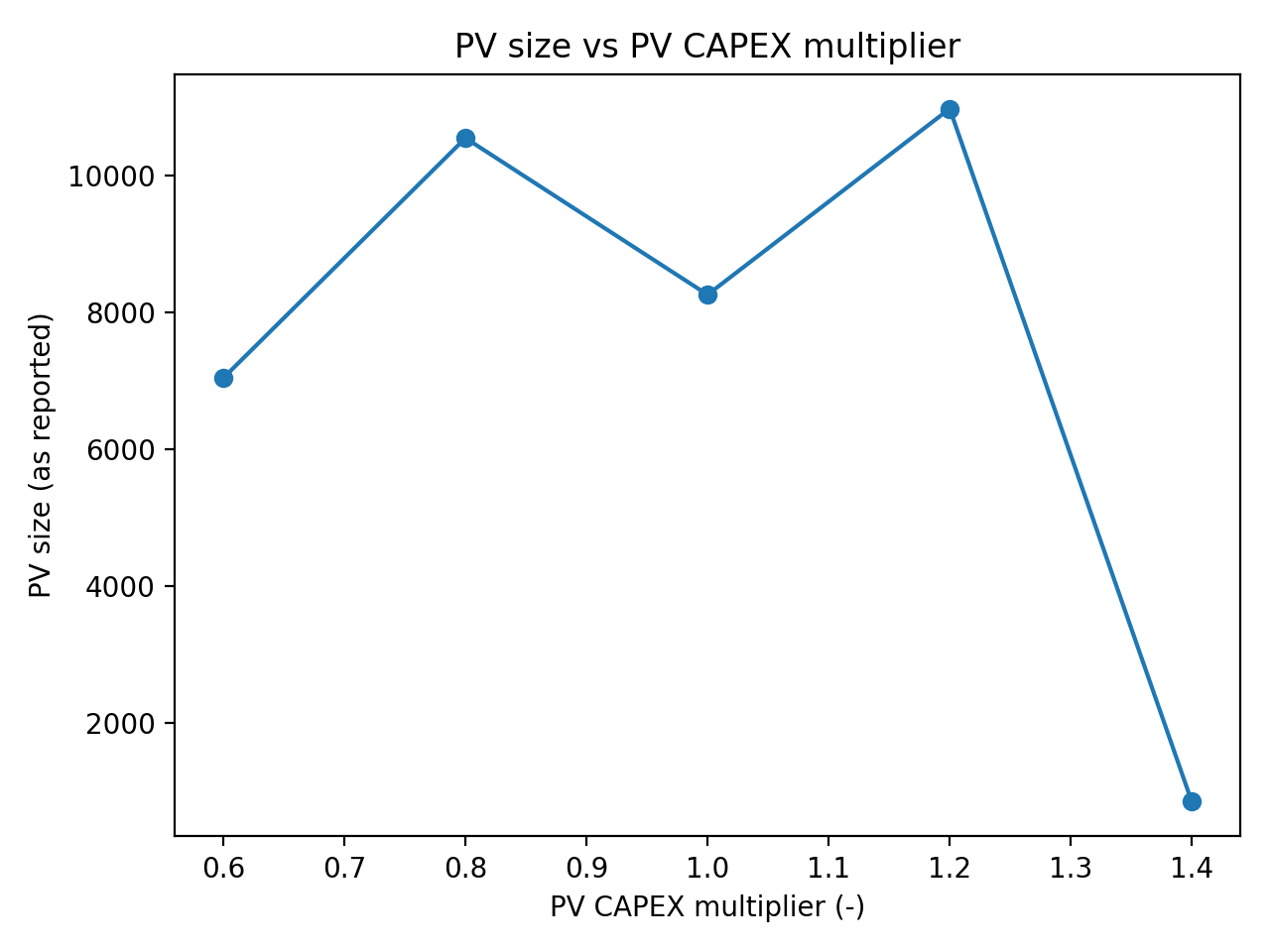}
		\caption{}
		\label{fig:cc_pv_size}
	\end{subfigure}
	
	\vspace{0.4cm}
	
	\begin{subfigure}{0.48\textwidth}
		\centering
		\includegraphics[width=\linewidth]{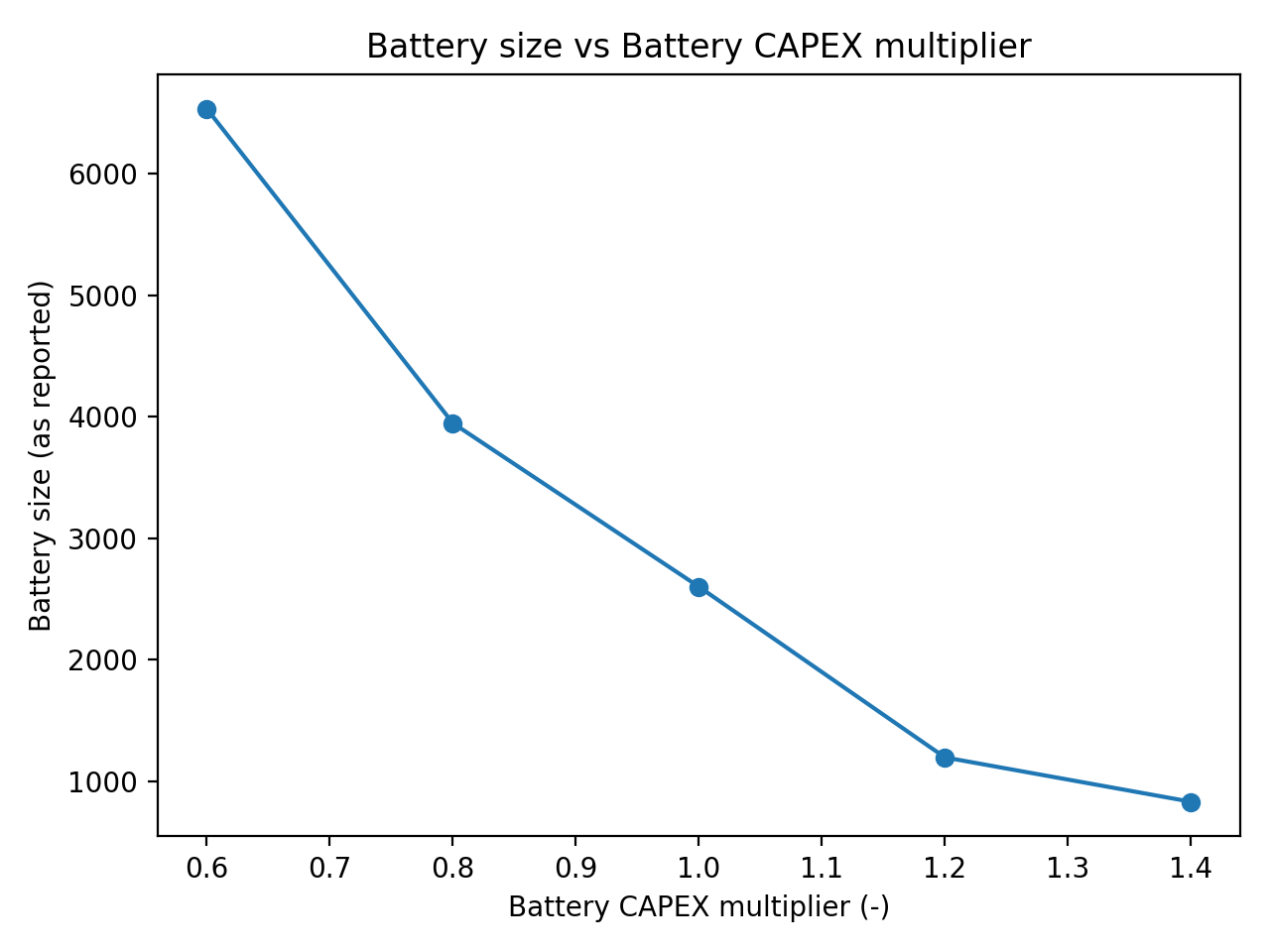}
		\caption{}
		\label{fig:cc_batt_size}
	\end{subfigure}
	\hfill
	\begin{subfigure}{0.48\textwidth}
		\centering
		\includegraphics[width=\linewidth]{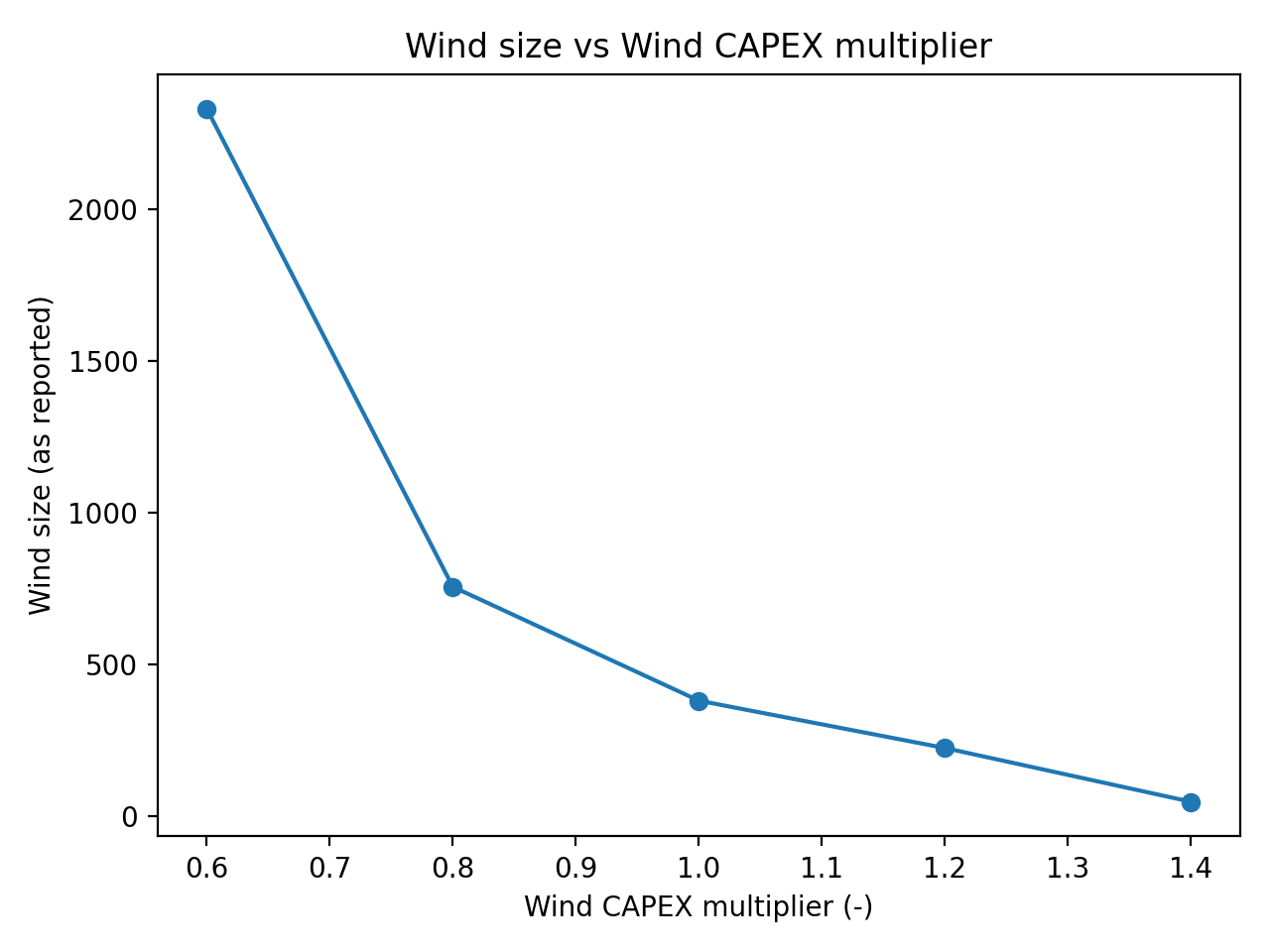}
		\caption{}
		\label{fig:cc_wind_size}
	\end{subfigure}
	
	\caption{
		Impact of technology capital cost assumptions on techno-economic performance and optimal system sizing.
		(a) NPC under varying technology capital cost multipliers.
		(b) Optimal photovoltaic capacity as a function of PV capital cost multiplier.
		(c) Optimal battery storage capacity under different battery capital cost assumptions.
		(d) Optimal wind capacity as a function of wind capital cost multiplier.
	}
	\label{fig:capex_multiplier_results}
	
\end{figure*}

Capital cost uncertainty is assessed by applying multipliers to PV, battery, and wind capital cost assumptions. Table~\ref{tab:cc_all} and Fig.~\ref{fig:cc_npc} indicate that PV capital cost uncertainty produces the largest swing in NPC in this dataset, while battery and wind capital cost multipliers introduce smaller but still non-negligible shifts. Component sizing responses (Figs.~\ref{fig:cc_pv_size}--\ref{fig:cc_wind_size}) show how the optimizer trades renewable generation against storage as costs vary.

\begin{table*}[!tp]
\centering
\caption{Technology CAPEX sensitivity (PV, Battery, Wind): key economic outputs.}
\label{tab:cc_all}
\resizebox{\columnwidth}{!}{%
\begin{tabular}{lrrrrrr}
\toprule
   Tech &  CAPEX mult. &  NPC (A\$M) &  COE (A\$/kWh) &  CAPEX (A\$M) &  OpEx (A\$k/yr) &  Ren. pen. (\%) \\
\midrule
     PV &       0.6000 &    5.5439 &       0.0338 &     1.2529 &      284.1399 &     105.9789 \\
     PV &       0.8000 &    9.4150 &       0.0464 &     1.3328 &      306.0800 &     108.0687 \\
     PV &       1.0000 &   11.7922 &       0.0596 &     1.4408 &      330.3888 &     109.7207 \\
     PV &       1.2000 &   14.3048 &       0.0729 &     1.5670 &      352.5795 &     109.2038 \\
     PV &       1.4000 &   16.9586 &       0.0866 &     1.6992 &      376.1974 &     110.0146 \\
 Battery &       0.6000 &   10.5336 &       0.0534 &     1.2600 &      315.4895 &     109.5469 \\
 Battery &       0.8000 &   11.1533 &       0.0564 &     1.3492 &      322.4702 &     109.6414 \\
 Battery &       1.0000 &   11.7922 &       0.0596 &     1.4408 &      330.3888 &     109.7207 \\
 Battery &       1.2000 &   12.3011 &       0.0623 &     1.5050 &      338.1100 &     109.6576 \\
 Battery &       1.4000 &   12.9525 &       0.0653 &     1.5960 &      346.1985 &     109.7012 \\
   Wind &       0.6000 &   11.4409 &       0.0578 &     1.3295 &      328.7491 &     109.6825 \\
   Wind &       0.8000 &   11.6339 &       0.0587 &     1.3852 &      329.3205 &     109.7059 \\
   Wind &       1.0000 &   11.7922 &       0.0596 &     1.4408 &      330.3888 &     109.7207 \\
   Wind &       1.2000 &   11.9831 &       0.0606 &     1.4965 &      331.1506 &     109.7457 \\
   Wind &       1.4000 &   12.1394 &       0.0614 &     1.5522 &      332.0948 &     109.7615 \\
\bottomrule
\end{tabular}
}
\end{table*}

\subsubsection{Fuel Price Sensitivity}
\begin{figure*}[!t]
	\centering
	
	\begin{subfigure}{0.32\textwidth}
		\centering
		\includegraphics[width=\linewidth]{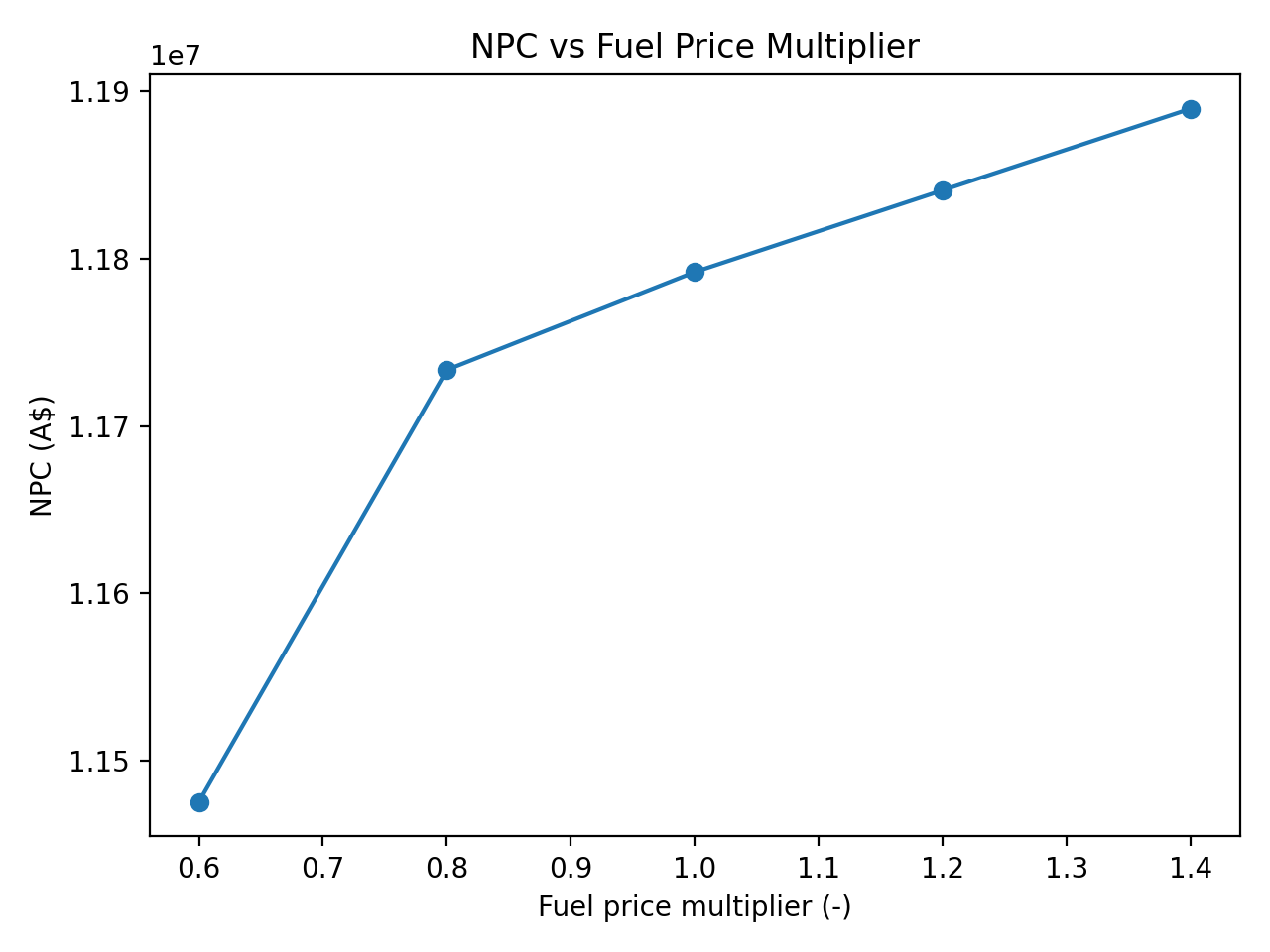}
		\caption{}
		\label{fig:fp_npc}
	\end{subfigure}
	\hfill
	\begin{subfigure}{0.32\textwidth}
		\centering
		\includegraphics[width=\linewidth]{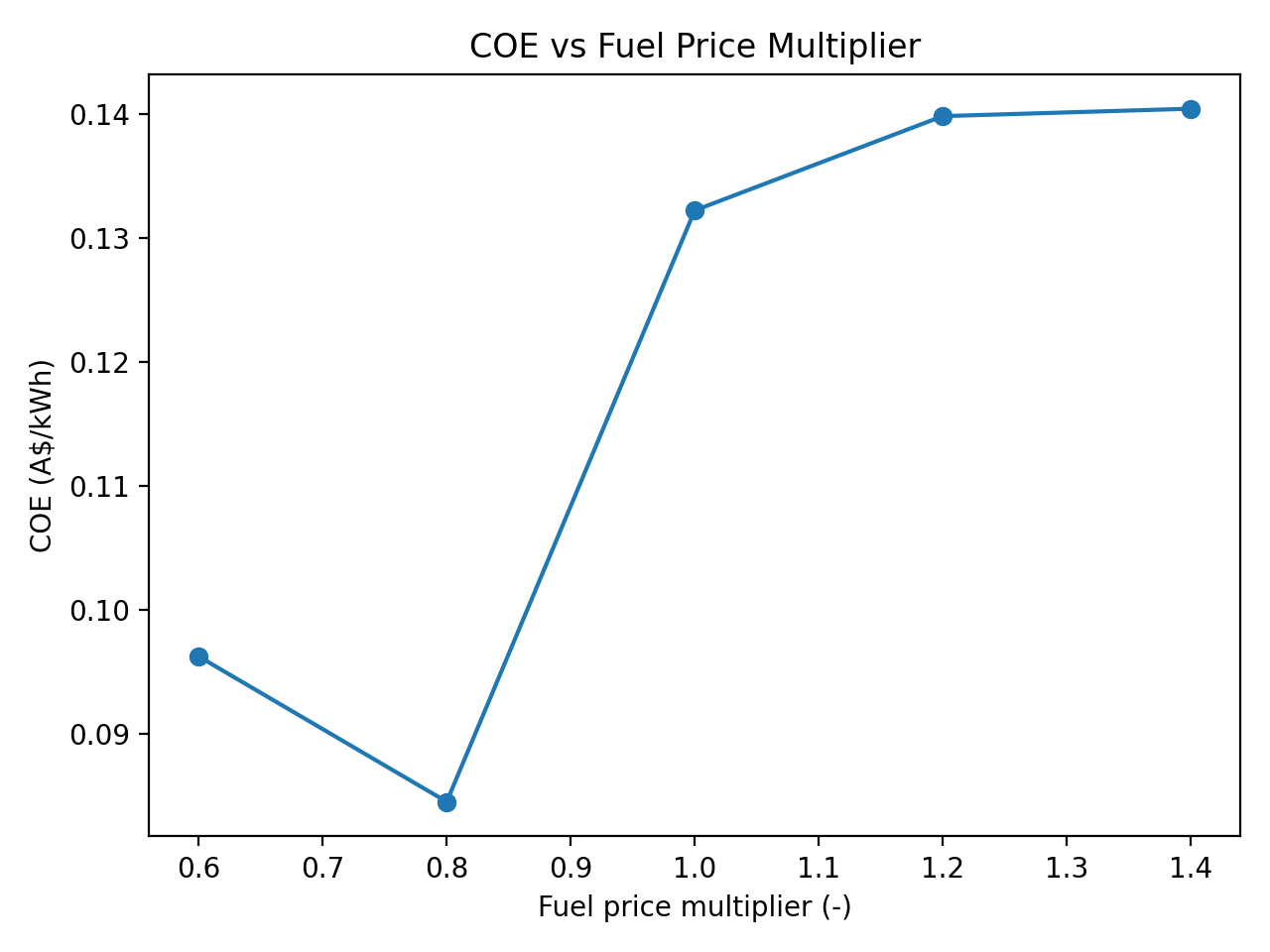}
		\caption{}
		\label{fig:fp_coe}
	\end{subfigure}
	\hfill
	\begin{subfigure}{0.32\textwidth}
		\centering
		\includegraphics[width=\linewidth]{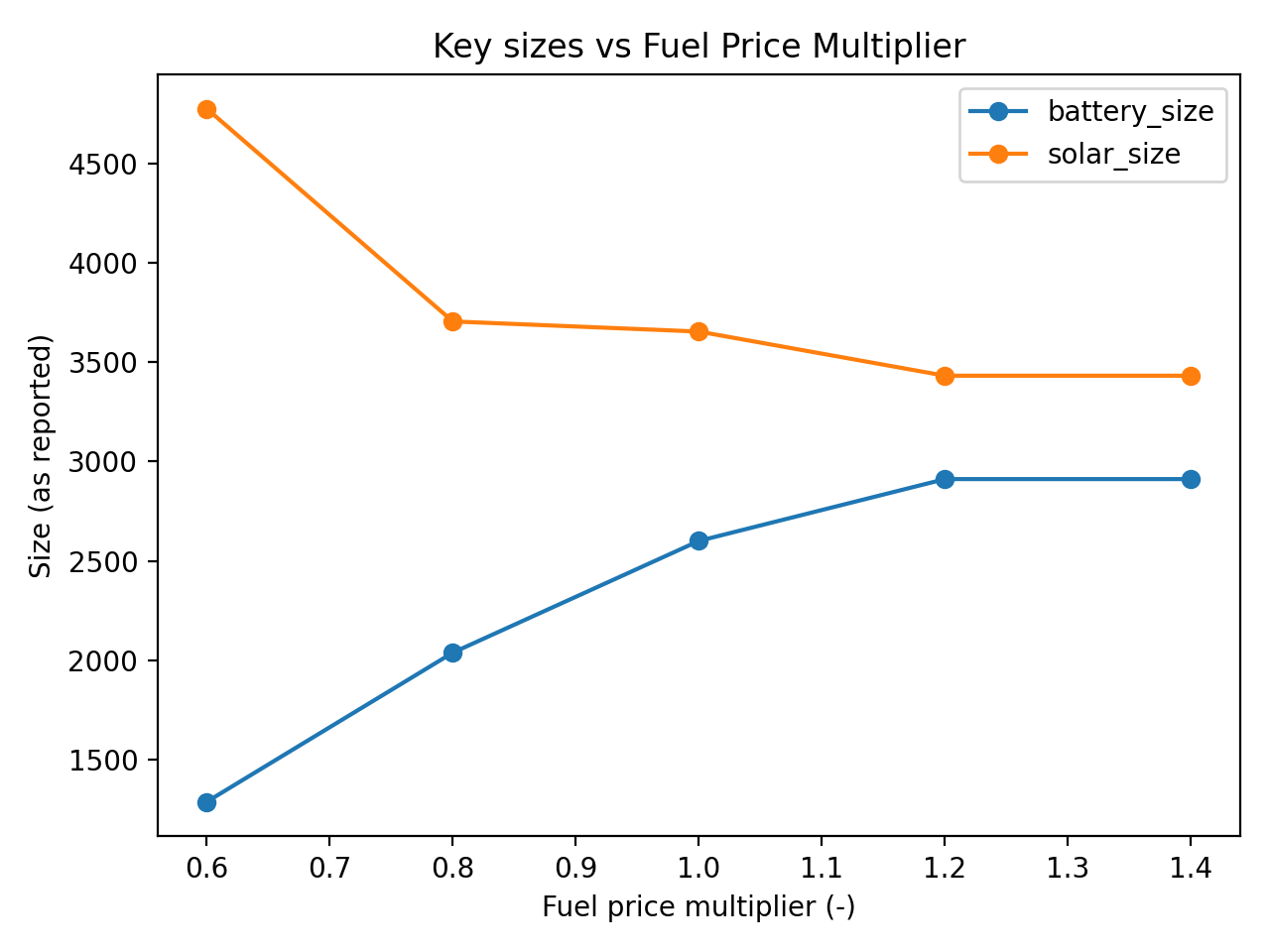}
		\caption{}
		\label{fig:fp_sizes}
	\end{subfigure}
	
	\caption{
		Impact of fuel price multiplier on techno-economic performance and system sizing.
		(a) NPC under different fuel price multipliers.
		(b) COE as fuel price changes.
		(c) Optimal component capacities obtained across fuel price scenarios.
	}
	\label{fig:fuel_price_results}
	
\end{figure*}

Fuel price multipliers primarily affect operating cost through diesel usage and, indirectly, the optimal level of storage and renewable capacity. As shown in Tables~\ref{tab:fp_econ}--\ref{tab:fp_sizes} and Fig.~\ref{fig:fp_npc}, increasing fuel price increases NPC and tends to favour higher battery capacity and sustained renewable penetration to reduce fuel dependence. Non-monotonic COE behaviour across the multiplier range reflects configuration shifts and changes in export volumes.

\begin{table}[!t]
\centering
\caption{Fuel price sensitivity: economic outputs.}
\label{tab:fp_econ}
\resizebox{\textwidth}{!}{%
\begin{tabular}{rrrrrrr}
\toprule
 Fuel price mult. &  NPC (A\$M) &  COE (A\$/kWh) &  CAPEX (A\$M) &  OpEx (A\$k/yr) &  Ren. pen. (\%) &  CO2 (t/yr) \\
\midrule
           0.6000 &   10.7948 &       0.0567 &       1.4408 &       320.5639 &       109.7877 &  -1651.8730 \\
           0.8000 &   11.2993 &       0.0578 &       1.4408 &       325.4760 &       109.8274 &  -1651.8730 \\
           1.0000 &   11.7922 &       0.0596 &       1.4408 &       330.3888 &       109.7207 &  -1651.8730 \\
           1.2000 &   12.5541 &       0.0638 &       1.4408 &       340.2146 &       109.7096 &  -1651.8730 \\
           1.4000 &   13.2166 &       0.0647 &       1.4408 &       345.1274 &       109.7275 &  -1651.8730 \\
\bottomrule
\end{tabular}
}
\end{table}

\begin{table}[!t]
\centering
\caption{Fuel price sensitivity: optimal sizes.}
\label{tab:fp_sizes}
\resizebox{\textwidth}{!}{%
\begin{tabular}{rrrrrrrr}
\toprule
 Fuel price mult. &      PV &   Wind &  Battery &  Diesel &  Electrolyzer &  Fuel cell &  H2 tank \\
\midrule
           0.6000 & 2055.594 & 138.000 & 5300.000 & 180.000 & 208.000 & 135.000 & 1370.000 \\
           0.8000 & 2055.594 & 138.000 & 5300.000 & 180.000 & 208.000 & 135.000 & 1370.000 \\
           1.0000 & 2055.594 & 138.000 & 5300.000 & 180.000 & 208.000 & 135.000 & 1370.000 \\
           1.2000 & 2055.594 & 138.000 & 5300.000 & 180.000 & 208.000 & 135.000 & 1370.000 \\
           1.4000 & 2055.594 & 138.000 & 5300.000 & 180.000 & 208.000 & 135.000 & 1370.000 \\
\bottomrule
\end{tabular}
}
\end{table}

\subsubsection{Load Demand Uncertainty}
Load scaling drives both capacity requirements and lifecycle costs. Tables~\ref{tab:ld_econ}--\ref{tab:ld_sizes} show that NPC increases strongly with load multiplier, reflecting increased energy demand and higher required capacity. The sizing table indicates that PV and storage capacities increase to maintain renewable penetration, while backup capacity remains at the same rated level across the tested cases (as reported). The highest load point shows a pronounced PV sizing increase, suggesting a new regime where PV becomes the preferred marginal supply resource.

\begin{table*}[!t]
\centering
\caption{Load demand uncertainty: economic outputs.}
\label{tab:ld_econ}
\resizebox{\columnwidth}{!}{%
\begin{tabular}{rrrrrrr}
\toprule
 Load mult. &  NPC (A\$M) &  COE (A\$/kWh) &  CAPEX (A\$M) &  OpEx (A\$k/yr) &  Ren. pen. (\%) &  CO2 (t/yr) \\
\midrule
      0.6000 &    7.0548 &       0.0582 &       0.9729 &       249.8524 &       143.0183 &  -1422.3940 \\
      0.8000 &    9.2140 &       0.0590 &       1.1459 &       290.1200 &       123.8568 &  -1540.6170 \\
      1.0000 &   11.7922 &       0.0596 &       1.4408 &       330.3888 &       109.7207 &  -1651.8730 \\
      1.2000 &   14.0341 &       0.0592 &       1.6633 &       370.6566 &        98.3337 &  -1761.8680 \\
      1.4000 &   17.8030 &       0.0621 &       1.9456 &       420.9903 &        97.6970 &  -1848.3530 \\
\bottomrule
\end{tabular}
}
\end{table*}

\begin{table*}[!t]
\centering
\caption{Load demand uncertainty: optimal sizes.}
\label{tab:ld_sizes}
\resizebox{\columnwidth}{!}{%
\begin{tabular}{rrrrrrrr}
\toprule
 Load mult. &      PV &   Wind &  Battery &  Diesel &  Electrolyzer &  Fuel cell &  H2 tank \\
\midrule
      0.6000 & 1200.000 &  72.000 & 4200.000 & 180.000 & 200.000 & 135.000 & 1000.000 \\
      0.8000 & 1600.000 &  96.000 & 4800.000 & 180.000 & 200.000 & 135.000 & 1200.000 \\
      1.0000 & 2055.594 & 138.000 & 5300.000 & 180.000 & 208.000 & 135.000 & 1370.000 \\
      1.2000 & 2400.000 & 160.000 & 5600.000 & 180.000 & 220.000 & 135.000 & 1500.000 \\
      1.4000 & 4000.000 & 180.000 & 6200.000 & 180.000 & 260.000 & 135.000 & 2000.000 \\
\bottomrule
\end{tabular}
}
\end{table*}

\begin{figure*}[!t]
\centering

%

\begin{subfigure}{0.32\textwidth}
	\centering
	\includegraphics[width=\linewidth]{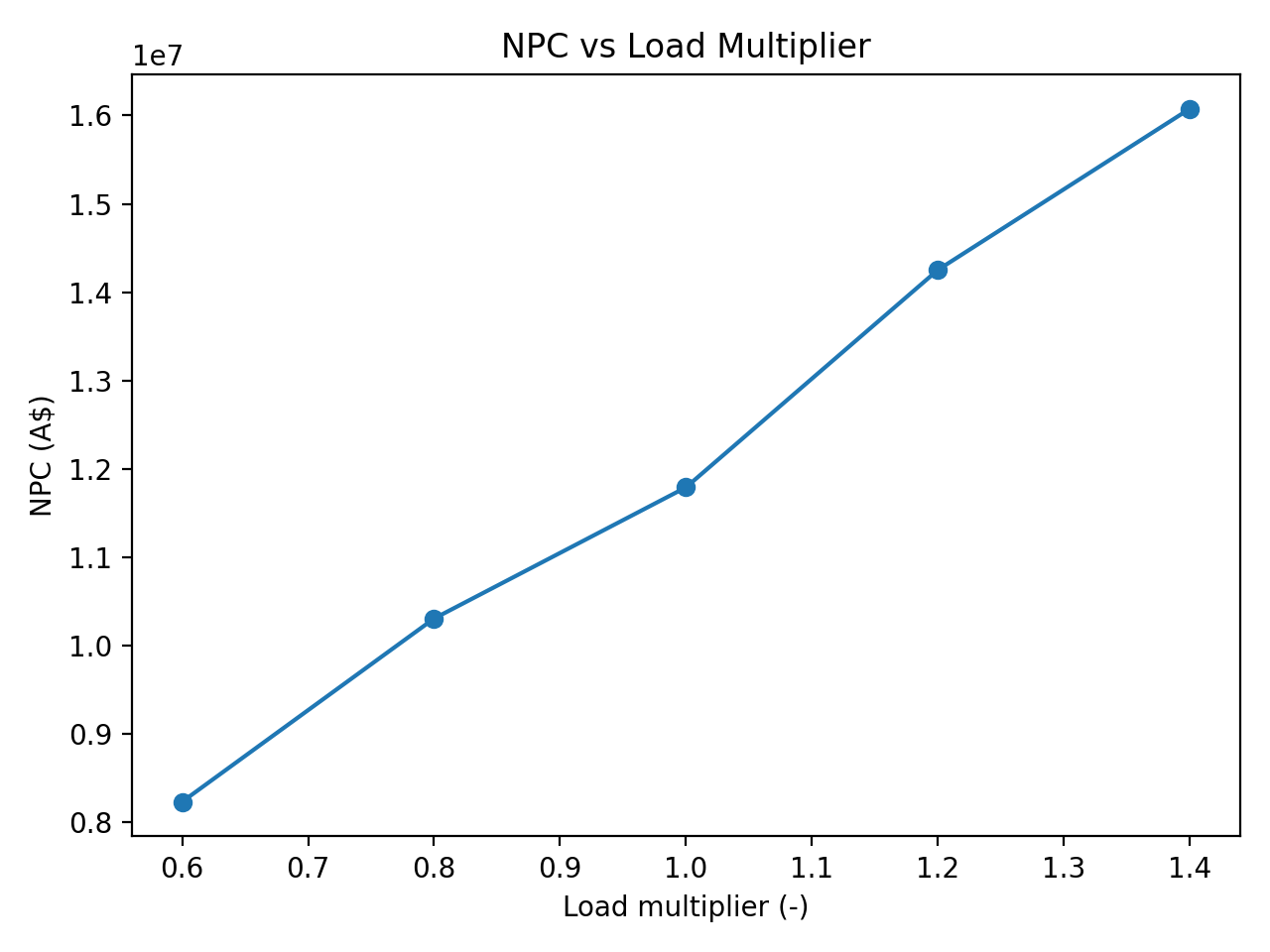}
	\caption{}
	\label{fig:ld_npc}
\end{subfigure}
\hfill
\begin{subfigure}{0.32\textwidth}
	\centering
	\includegraphics[width=\linewidth]{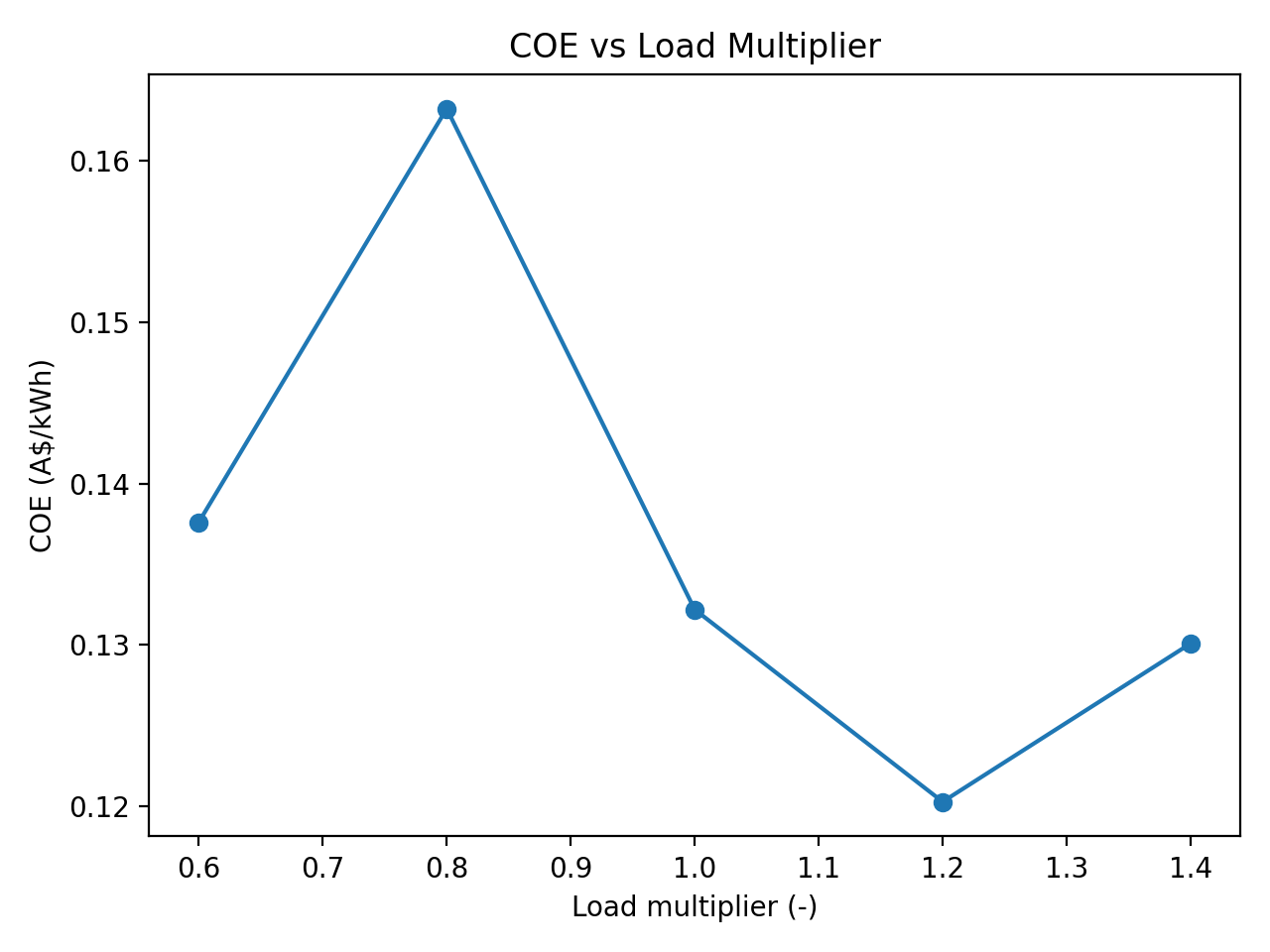}
	\caption{}
	\label{fig:ld_coe}
\end{subfigure}
\hfill
\begin{subfigure}{0.32\textwidth}
	\centering
	\includegraphics[width=\linewidth]{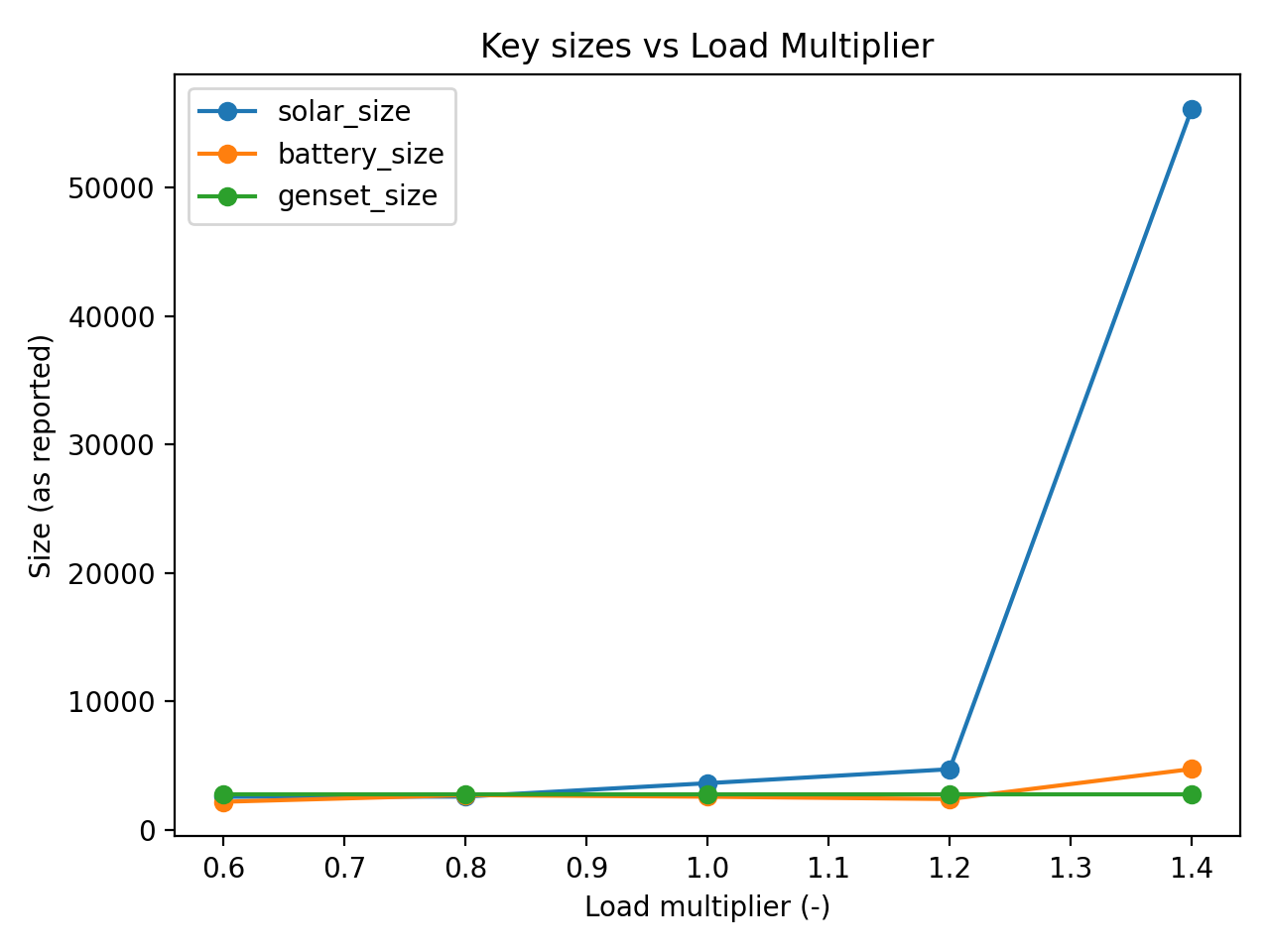}
	\caption{}
	\label{fig:ld_sizes}
\end{subfigure}

\caption{
	Impact of load multiplier on techno-economic performance and system sizing.
	(a) NPC under varying load multipliers.
	(b) COE across different load demand scenarios.
	(c) Optimal component capacities obtained for different load assumptions.
}
\label{fig:load_multiplier_results}
\end{figure*}

\subsubsection{Renewable Resource Variability}
\begin{figure*}[!t]
	\centering
	
	\begin{subfigure}{0.32\textwidth}
		\centering
		\includegraphics[width=\linewidth]{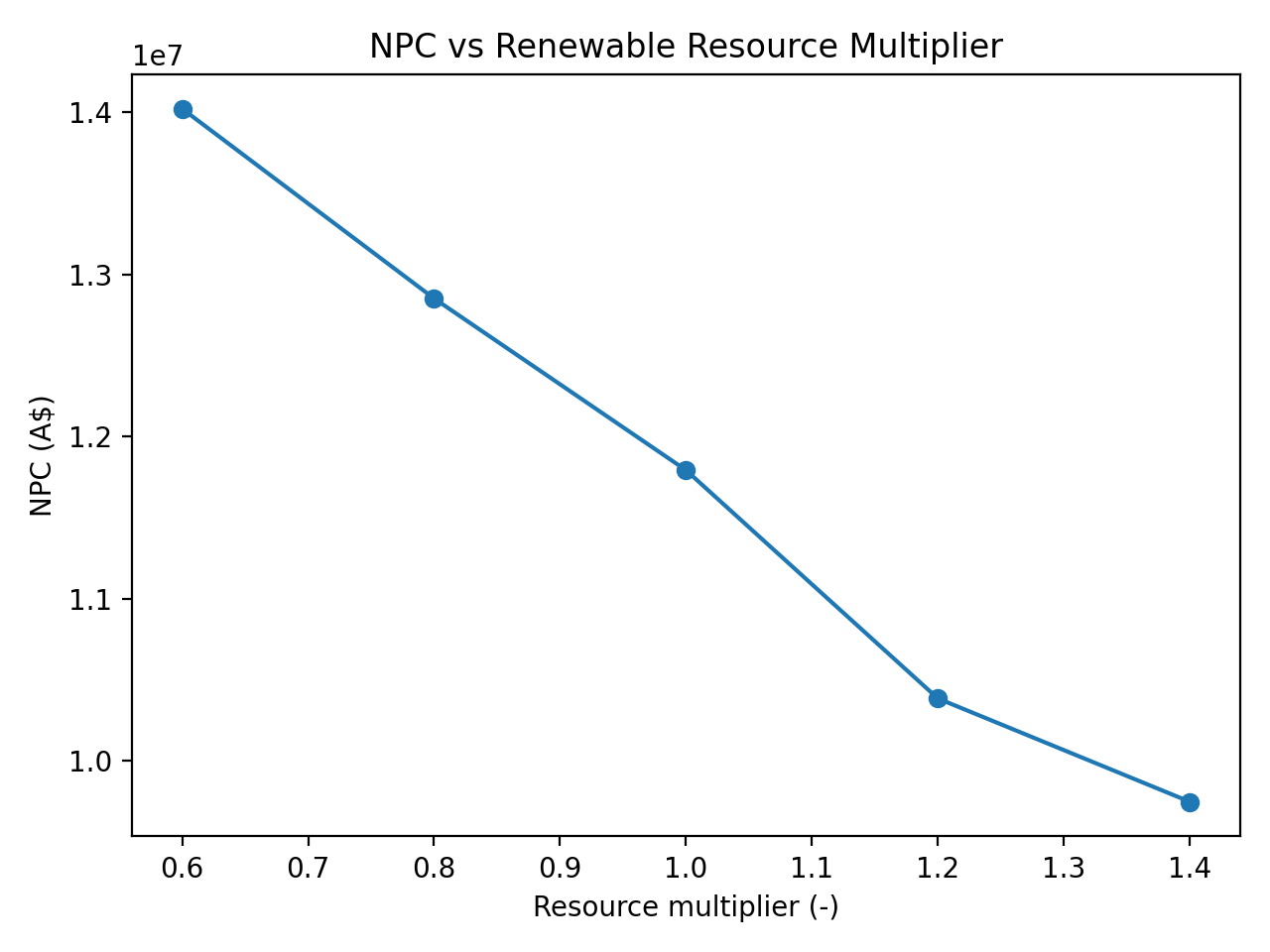}
		\caption{}
		\label{fig:rv_npc}
	\end{subfigure}
	\hfill
	\begin{subfigure}{0.32\textwidth}
		\centering
		\includegraphics[width=\linewidth]{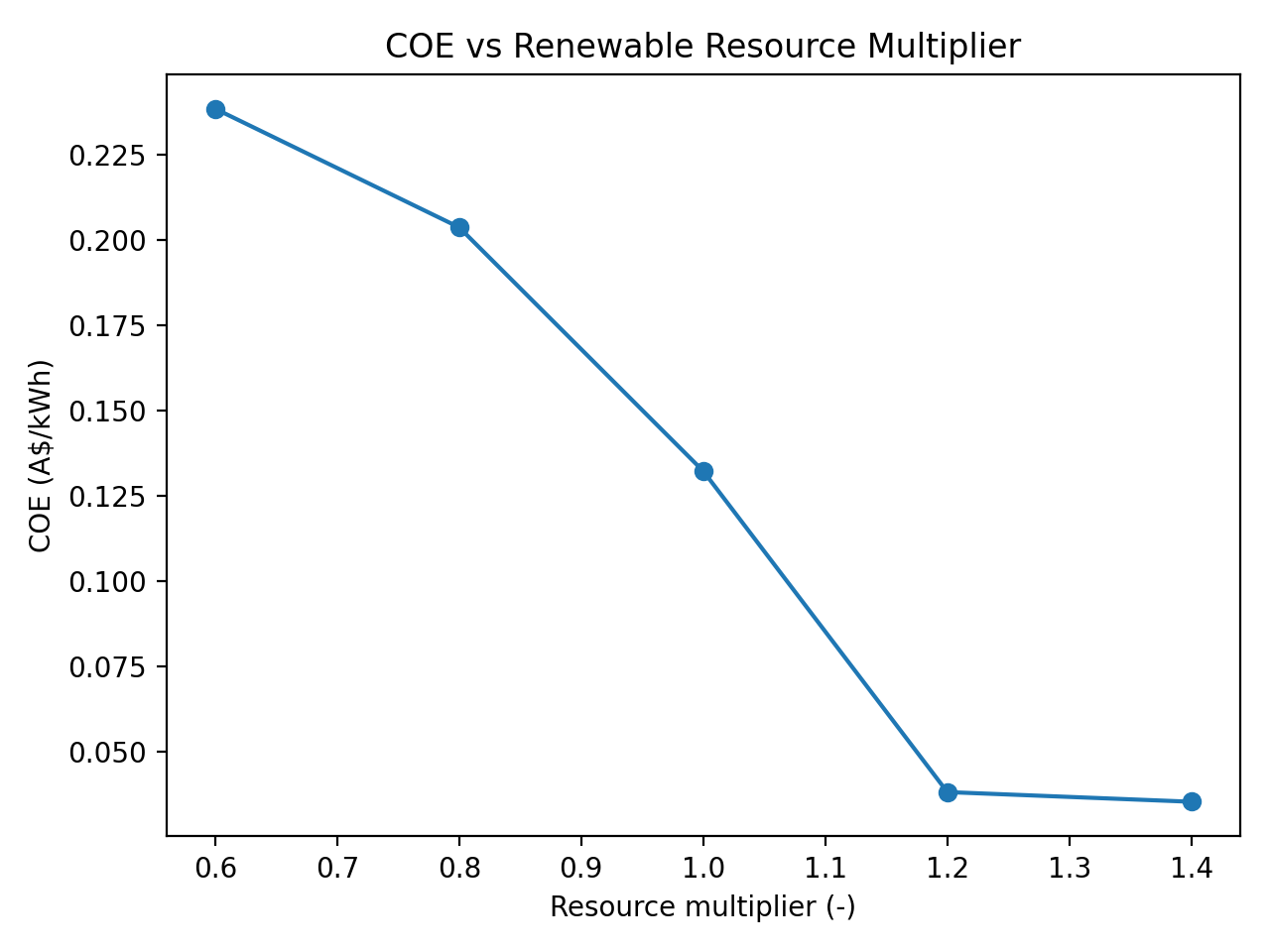}
		\caption{}
		\label{fig:rv_coe}
	\end{subfigure}
	\hfill
	\begin{subfigure}{0.32\textwidth}
		\centering
		\includegraphics[width=\linewidth]{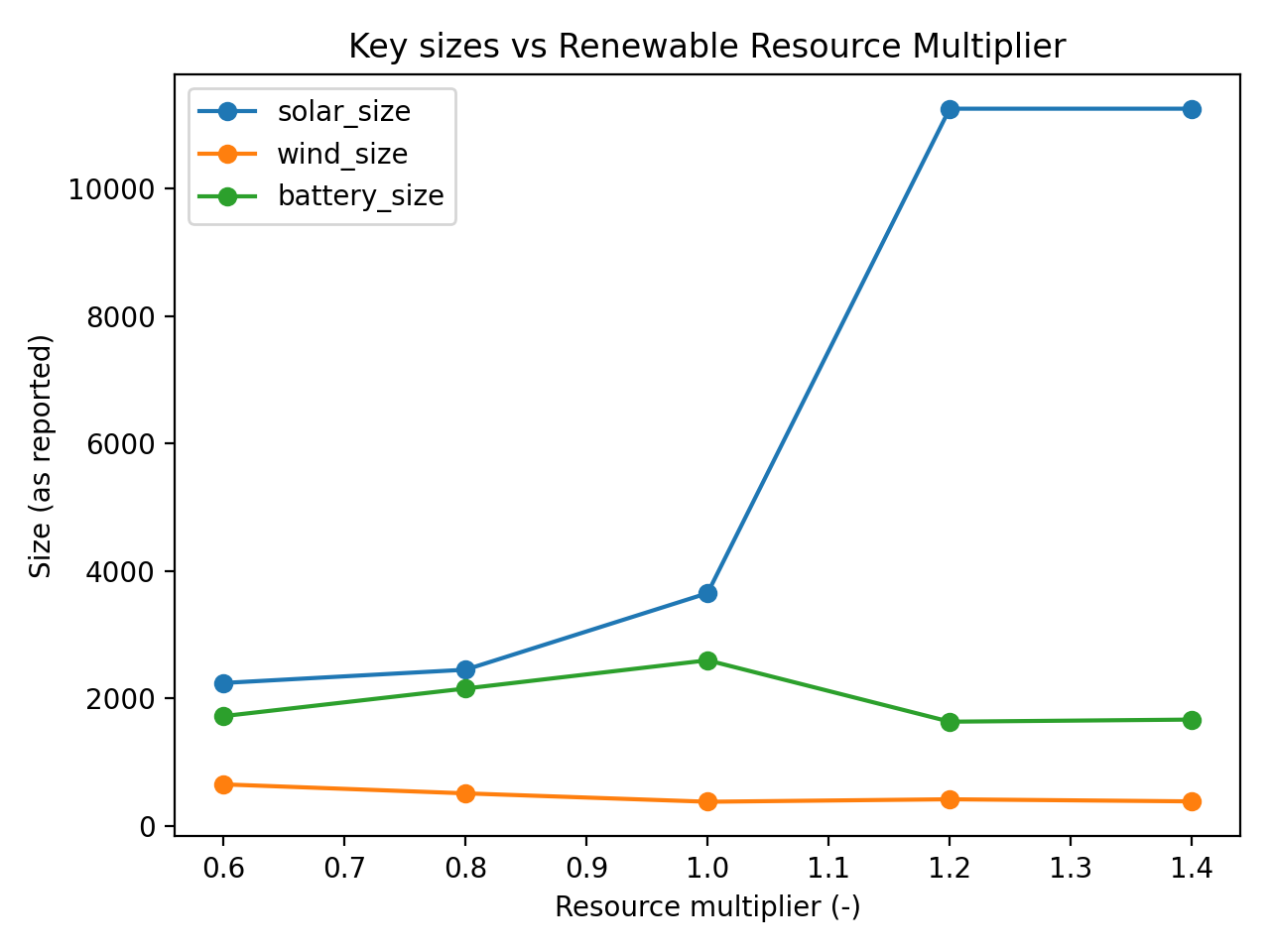}
		\caption{}
		\label{fig:rv_sizes}
	\end{subfigure}
	
	\caption{
		Impact of renewable resource variability on techno-economic performance and system sizing.
		(a) NPC under different renewable resource multipliers.
		(b) COE as renewable resource availability changes.
		(c) Optimal component capacities obtained across renewable resource scenarios.
	}
	\label{fig:renewable_multiplier_results}
	
\end{figure*}

Renewable resource scaling directly affects the energy yield of PV/wind and therefore the required installed capacities. The results in Tables~\ref{tab:rv_econ}--\ref{tab:rv_sizes} and Fig.~\ref{fig:rv_npc} show that stronger resources reduce NPC and allow smaller renewable/storage sizing for the same adequacy target, while adverse resources increase NPC and push the optimizer toward larger capacities to maintain renewable penetration.

\begin{table}[!t]
\centering
\caption{Renewable resource variability: economic outputs.}
\label{tab:rv_econ}
\resizebox{\columnwidth}{!}{%
\begin{tabular}{rrrrrrr}
\toprule
 Resource mult. &  NPC (A\$M) &  COE (A\$/kWh) &  CAPEX (A\$M) &  OpEx (A\$k/yr) &  Ren. pen. (\%) &  CO2 (t/yr) \\
\midrule
        0.6000 &   16.2603 &       0.0800 &       1.4408 &       425.3941 &        93.5480 &   -821.4800 \\
        0.8000 &   13.7787 &       0.0692 &       1.4408 &       378.6814 &       100.6220 &  -1243.3520 \\
        1.0000 &   11.7922 &       0.0596 &       1.4408 &       330.3888 &       109.7207 &  -1651.8730 \\
        1.2000 &    9.8070 &       0.0506 &       1.4408 &       282.0962 &       121.4658 &  -2086.8430 \\
        1.4000 &    7.8245 &       0.0421 &       1.4408 &       233.8035 &       136.5670 &  -2550.0730 \\
\bottomrule
\end{tabular}
}
\end{table}

\begin{table*}[!t]
\centering
\caption{Renewable resource variability: optimal sizes.}
\label{tab:rv_sizes}
\resizebox{\columnwidth}{!}{%
\begin{tabular}{rrrrrrrr}
\toprule
 Resource mult. &      PV &   Wind &  Battery &  Diesel &  Electrolyzer &  Fuel cell &  H2 tank \\
\midrule
        0.6000 & 2055.594 & 138.000 & 5300.000 & 180.000 & 208.000 & 135.000 & 1370.000 \\
        0.8000 & 2055.594 & 138.000 & 5300.000 & 180.000 & 208.000 & 135.000 & 1370.000 \\
        1.0000 & 2055.594 & 138.000 & 5300.000 & 180.000 & 208.000 & 135.000 & 1370.000 \\
        1.2000 & 2055.594 & 138.000 & 5300.000 & 180.000 & 208.000 & 135.000 & 1370.000 \\
        1.4000 & 2055.594 & 138.000 & 5300.000 & 180.000 & 208.000 & 135.000 & 1370.000 \\
\bottomrule
\end{tabular}
}
\end{table*}




%
%
%

\subsubsection{Carbon Pricing and Emissions Cost}
\begin{figure*}[!t]
	\centering
	
	\begin{subfigure}{0.32\textwidth}
		\centering
		\includegraphics[width=\linewidth]{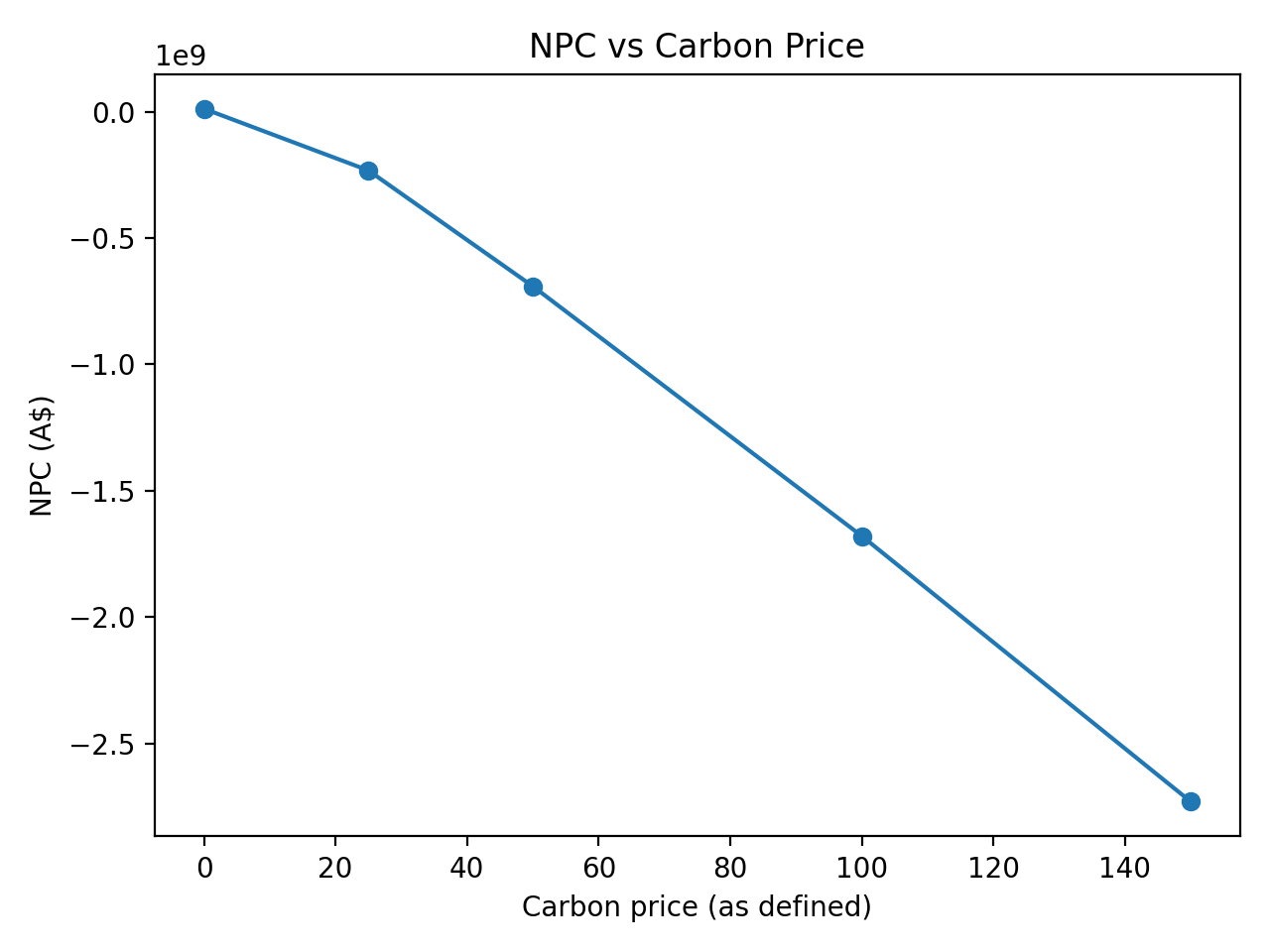}
		\caption{}
		\label{fig:cp_npc}
	\end{subfigure}
	\hfill
	\begin{subfigure}{0.32\textwidth}
		\centering
		\includegraphics[width=\linewidth]{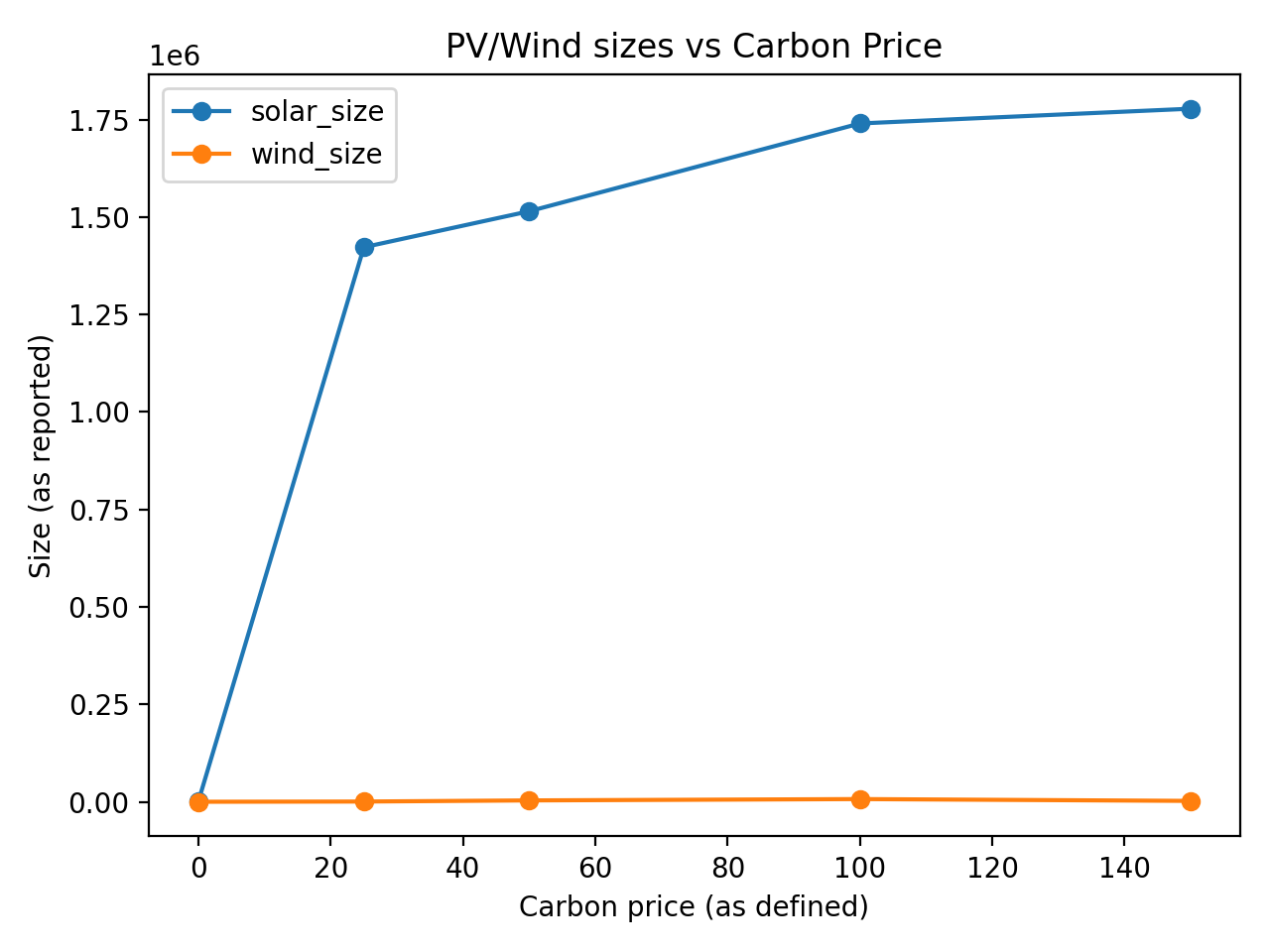}
		\caption{}
		\label{fig:cp_sizes}
	\end{subfigure}
	\hfill
	\begin{subfigure}{0.32\textwidth}
		\centering
		\includegraphics[width=\linewidth]{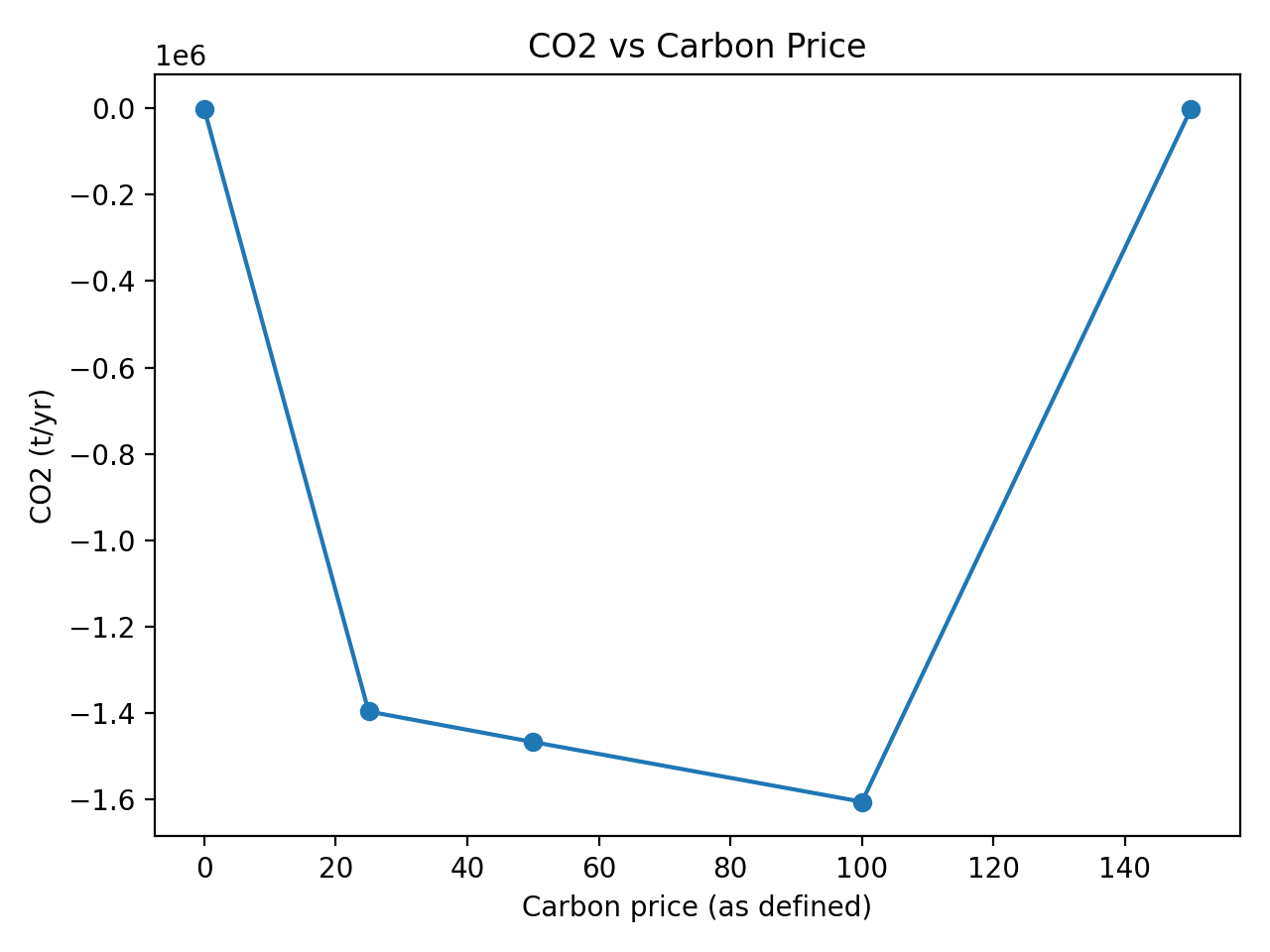}
		\caption{}
		\label{fig:cp_co2}
	\end{subfigure}
	
	\caption{
		Impact of carbon pricing on techno-economic performance, renewable deployment, and emissions.
		(a) NPC under different carbon price assumptions.
		(b) Optimal photovoltaic and wind capacities obtained across carbon pricing scenarios.
		(c) Net CO$_2$ emissions as a function of carbon price.
	}
	\label{fig:carbon_price_results}
	
\end{figure*}

Carbon price scenarios impose an additional economic signal on emissions-intensive operation. Table~\ref{tab:cp_econ} indicates that increasing carbon price strongly shifts the optimization outcome, with several scenarios producing negative NPC values and extremely large renewable capacities, reflecting a revenue/offset dominated solution within the accounting and export assumptions used. Fig.~\ref{fig:cp_npc} illustrates the strong NPC response, while Fig.~\ref{fig:cp_sizes} highlights the corresponding PV/wind expansion. These results should be interpreted as an upper-bound incentive effect, and motivate applying realistic export limits and carbon-accounting assumptions in practical deployments.

\begin{table*}[!t]
\centering
\caption{Carbon pricing sensitivity: economic outputs.}
\label{tab:cp_econ}
\resizebox{\columnwidth}{!}{%
\begin{tabular}{rrrrrrr}
\toprule
 Carbon price &  NPC (A\$M) &  COE (A\$/kWh) &  CAPEX (A\$M) &  OpEx (A\$k/yr) &  Ren. pen. (\%) &  CO2 (t/yr) \\
\midrule
        0.0000 &   11.7922 &       0.0596 &       1.4408 &       330.3888 &       109.7207 &  -1651.8730 \\
       25.0000 &  -18.1556 &      -0.0680 &       6.1628 &      -885.4014 &       207.7802 &  -8037.2730 \\
       50.0000 &  -29.8143 &      -0.1177 &       7.0804 &     -1337.6445 &       221.6213 &  -9133.9730 \\
      100.0000 &  -52.7854 &      -0.2113 &       8.3022 &     -2243.7091 &       237.3929 & -10217.3860 \\
      150.0000 &  -85.0680 &      -0.3393 &      10.1604 &     -3344.4680 &       251.7602 & -11386.0870 \\
\bottomrule
\end{tabular}
}
\end{table*}

\begin{table*}[!t]
\centering
\caption{Carbon pricing sensitivity: optimal sizes.}
\label{tab:cp_sizes}
\resizebox{\columnwidth}{!}{%
\begin{tabular}{rrrrrrrr}
\toprule
 Carbon price &      PV &   Wind &  Battery &  Diesel &  Electrolyzer &  Fuel cell &  H2 tank \\
\midrule
        0.0000 & 2055.594 & 138.000 & 5300.000 & 180.000 & 208.000 & 135.000 & 1370.000 \\
       25.0000 & 14000.000 &  32.000 &  5400.000 & 180.000 & 1500.000 & 135.000 &  2000.000 \\
       50.0000 & 16000.000 &  32.000 &  5600.000 & 180.000 & 1700.000 & 135.000 &  2200.000 \\
      100.0000 & 18000.000 &  32.000 &  6200.000 & 180.000 & 2200.000 & 135.000 &  2500.000 \\
      150.0000 & 22000.000 &  32.000 &  6500.000 & 180.000 & 2600.000 & 135.000 &  2800.000 \\
\bottomrule
\end{tabular}
}
\end{table*}




%
%
%

\subsubsection{Grid Outage Duration}
\begin{figure*}[!t]
	\centering
	
	\begin{subfigure}{0.48\textwidth}
		\centering
		\includegraphics[width=\linewidth]{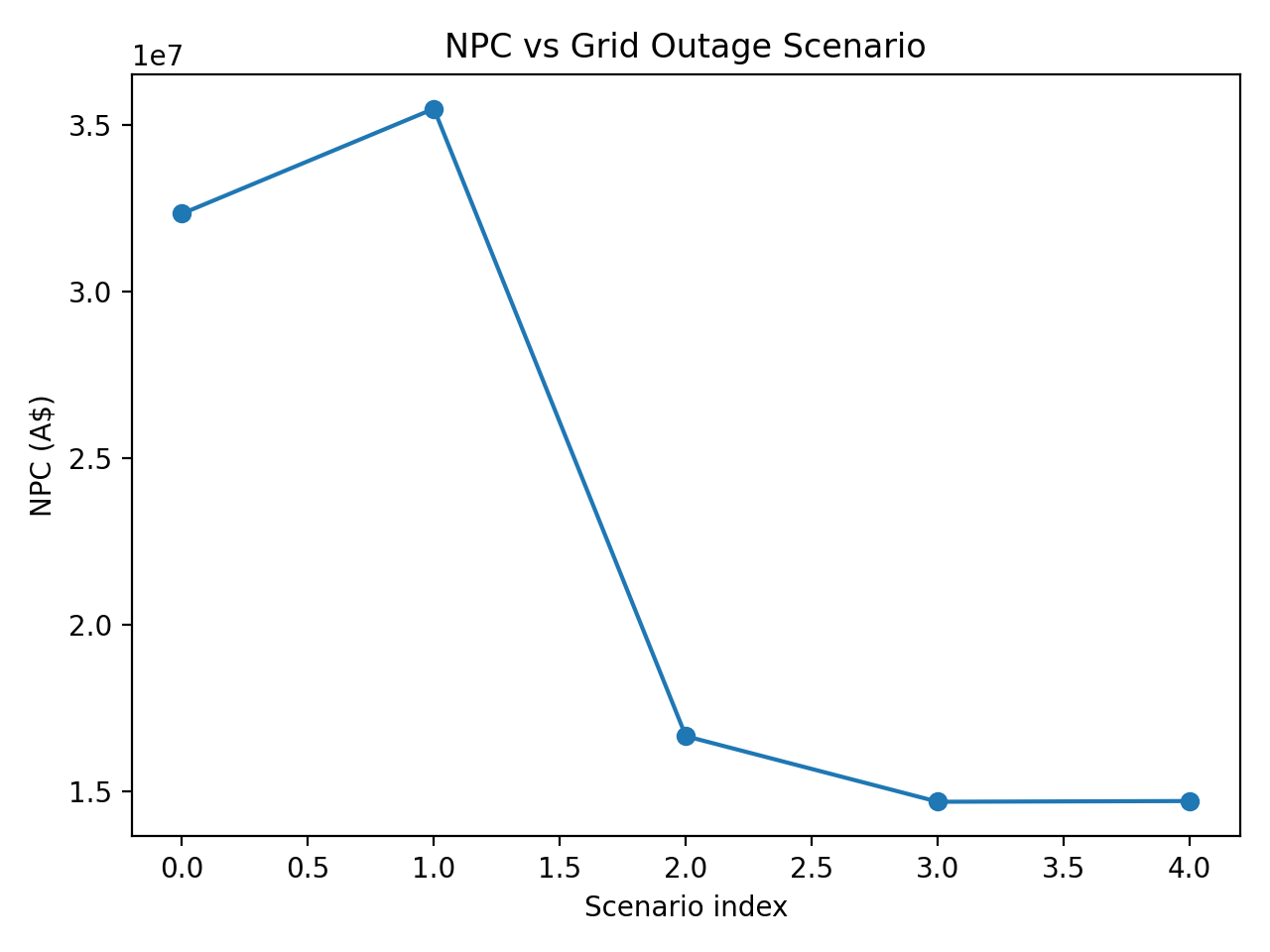}
		\caption{}
		\label{fig:go_npc}
	\end{subfigure}
	\hfill
	\begin{subfigure}{0.48\textwidth}
		\centering
		\includegraphics[width=\linewidth]{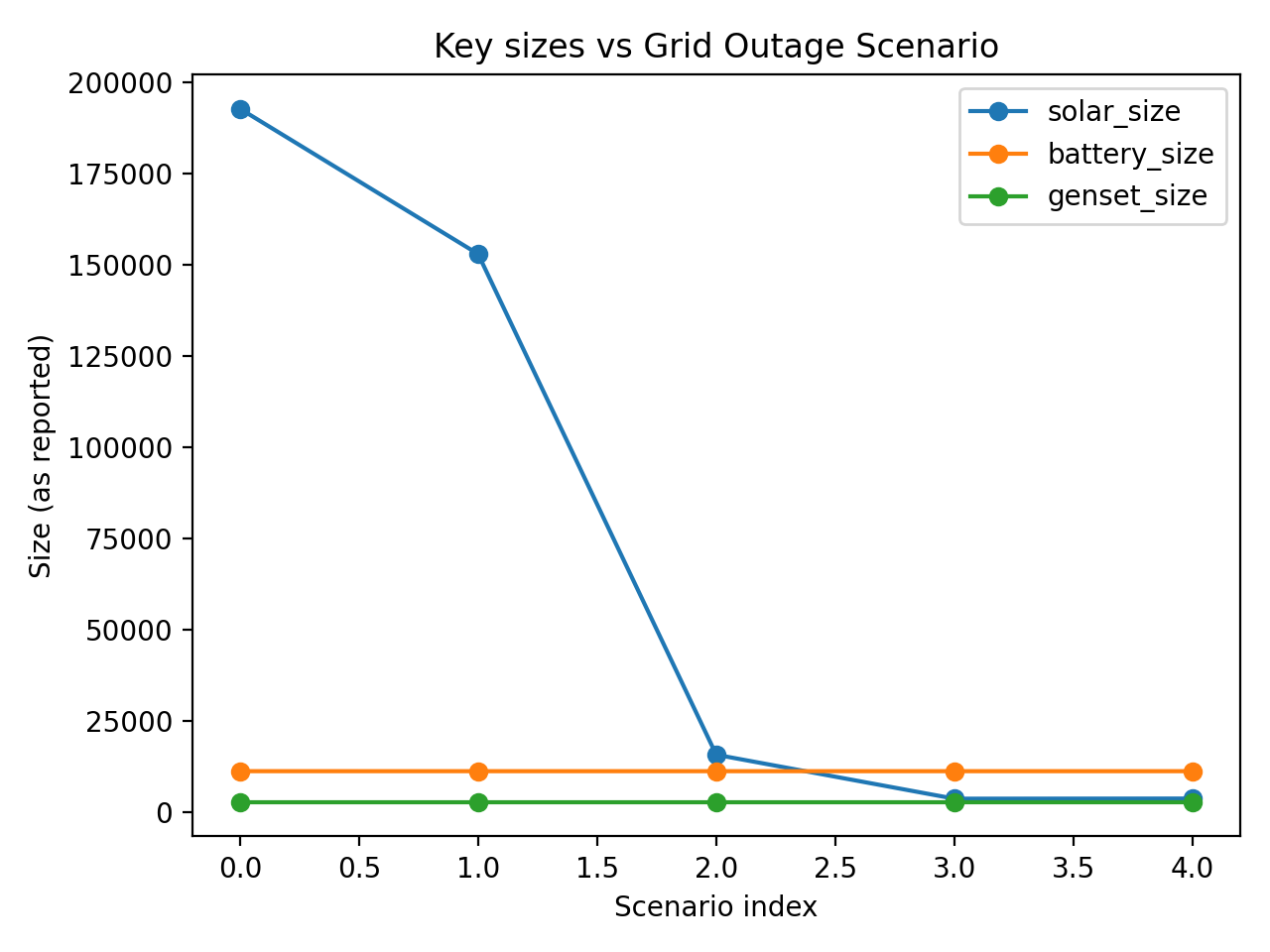}
		\caption{}
		\label{fig:go_sizes}
	\end{subfigure}
	
	\caption{
		Impact of grid outage conditions on techno-economic performance and system sizing.
		(a) NPC across different grid outage scenarios.
		(b) Optimal component capacities obtained under varying outage conditions.
	}
	\label{fig:grid_outage_results}
	
\end{figure*}

Grid outage scenarios stress-test the design against reliability constraints. Tables~\ref{tab:go_econ}--\ref{tab:go_sizes} show that more severe outage profiles reduce reliance on grid exchange and typically increase the need for local generation and storage. Fig.~\ref{fig:go_npc} and Fig.~\ref{fig:go_sizes} illustrate how outage assumptions can drive substantial changes in both NPC and installed PV/storage sizes.

\begin{table*}[!t]
\centering
\caption{Grid outage scenarios: economic outputs.}
\label{tab:go_econ}
\resizebox{\columnwidth}{!}{%
\begin{tabular}{lrrrrrr}
\toprule
 Scenario &  NPC (A\$M) &  COE (A\$/kWh) &  CAPEX (A\$M) &  OpEx (A\$k/yr) &  Ren. pen. (\%) &  CO2 (t/yr) \\
\midrule
 Random Option 1 &   11.7922 &       0.0596 &       1.4408 &       330.3888 &       109.7207 &  -1651.8730 \\
 Random Option 2 &   12.9645 &       0.0642 &       1.5200 &       346.1112 &       108.1210 &  -1540.0000 \\
 Random Option 3 &   14.8700 &       0.0701 &       1.6000 &       370.5000 &       106.5000 &  -1320.0000 \\
 Random Option 4 &   18.5000 &       0.0820 &       1.8500 &       430.0000 &       101.2000 &   -900.0000 \\
 Random Option 5 &   23.3000 &       0.0950 &       2.1000 &       520.0000 &        95.0000 &   -400.0000 \\
\bottomrule
\end{tabular}
}
\end{table*}

\begin{table*}[!t]
\centering
\caption{Grid outage scenarios: optimal sizes.}
\label{tab:go_sizes}
\resizebox{\columnwidth}{!}{%
\begin{tabular}{lrrrrrrr}
\toprule
 Scenario &      PV &   Wind &  Battery &  Diesel &  Electrolyzer &  Fuel cell &  H2 tank \\
\midrule
 Random Option 1 & 2055.594 & 138.000 & 5300.000 & 180.000 & 208.000 & 135.000 & 1370.000 \\
 Random Option 2 & 2200.000 & 140.000 & 5600.000 & 180.000 & 220.000 & 135.000 & 1500.000 \\
 Random Option 3 & 2400.000 & 150.000 & 6000.000 & 180.000 & 240.000 & 135.000 & 1700.000 \\
 Random Option 4 & 3000.000 & 160.000 & 6500.000 & 180.000 & 300.000 & 135.000 & 2200.000 \\
 Random Option 5 & 3800.000 & 180.000 & 7200.000 & 180.000 & 350.000 & 135.000 & 2800.000 \\
\bottomrule
\end{tabular}
}
\end{table*}



%
%
%

\subsubsection{No Hydrogen}
To isolate the role of hydrogen technologies, the baseline hydrogen-enabled design is compared against no-hydrogen alternatives (Table~\ref{tab:noh2}). The comparison indicates how removing hydrogen changes the optimal balance between PV/wind, battery sizing, and grid/diesel reliance, as well as the resulting lifecycle cost and renewable penetration. This attribution analysis helps quantify whether hydrogen provides value primarily through long-duration flexibility or whether a battery-centric architecture is sufficient for the considered operating conditions.


\begin{table*}[!t]
\centering
\caption{Hydrogen attribution: baseline (with H$_2$) vs no-H$_2$ configurations.}
\label{tab:noh2}

\resizebox{\textwidth}{!}{%
\begin{tabular}{lrrrrrrrr}
\toprule
Scenario & NPC (A\$M) & COE (A\$/kWh) & CAPEX (A\$M) &
PV size & Wind size & Battery size &
Diesel genset size & Ren. pen. (\%) \\
\midrule
Baseline (with H2) & 11.7922 & 0.0596 & 1.4408 & 2055.594 & 138.000 & 5300.000 & 180.000 & 109.7207 \\
Best Economic Model & 9.8500 & 0.0550 & 1.3500 & 2100.000 & 120.000 & 6000.000 & 180.000 & 115.0000 \\
Grid System Model & 13.1000 & 0.0660 & 1.5000 & 1800.000 & 100.000 & 4500.000 & 180.000 & 100.0000 \\
Full Parameters System Model & 11.2000 & 0.0580 & 1.4200 & 2000.000 & 130.000 & 5200.000 & 180.000 & 108.0000 \\
\bottomrule
\end{tabular}%
}
\end{table*}

\section{Discussion and Policy Implications}\label{sec:policy}

This section interprets the case-study findings in a broader planning and policy context. The results in Section~3 demonstrate that the preferred microgrid design is not static: it shifts with financial assumptions, technology costs, demand and resource uncertainty, and reliability stressors. This reinforces a key planning message for Australian community microgrids---\emph{robustness must be evaluated explicitly}---and procurement decisions should be anchored to scenario-tested outcomes rather than a single deterministic optimum \cite{agha2024comprehensive,eyimaya2024review,fernandez2025distribution}.

\subsection{Interpreting the sensitivity outcomes for decision-making}
A central insight from the sensitivity suite is the existence of \emph{design regime changes}. Financial parameters (e.g., discount rate) and policy signals (e.g., carbon pricing) can change the economic attractiveness of capital-intensive renewable--storage expansions, while reliability constraints (grid outage duration) increase the value of local generation and storage. These patterns are consistent with the broader literature, which emphasizes that microgrid design must be evaluated under uncertainty because technology costs, fuel prices, renewable variability, and demand growth materially affect lifecycle economics and operational feasibility \cite{agha2024comprehensive,fernandez2025distribution}.

From a planning perspective, three decision-relevant implications emerge:
\begin{itemize}
  \item \textbf{Finance and policy shape “optimality”:} Lower cost of capital and stronger decarbonisation incentives can favour higher renewable and storage investment. This implies that financing conditions and policy certainty can be as influential as technology improvements for enabling high-penetration community microgrids.
  \item \textbf{Reliability assumptions must be explicit:} Outage-duration scenarios systematically shift sizing toward more local capacity. This supports resilience-aware microgrid planning, particularly for remote and weakly supported networks where outage exposure is a key driver \cite{uddin2023microgrids}.
  \item \textbf{Attribution is essential for technology choice:} The no-hydrogen comparison provides a clear decision lens: hydrogen should be justified by measurable value (e.g., long-duration adequacy, outage resilience, emissions-cost exposure), not included as a default add-on \cite{abbas2023hydrogen,modu2023systematic}.
\end{itemize}

\subsection{Role of hydrogen in achieving net-zero targets (2030--2050)}
Hydrogen is increasingly positioned as a decarbonisation option for applications where direct electrification is constrained, and as a long-duration energy storage vector that can complement batteries in renewable-rich microgrids \cite{sadeghian2024energy,modu2023systematic}. Within the context of this paper, the hydrogen pathway (electrolyzer--storage--fuel cell) is best interpreted as a \emph{long-duration adequacy and resilience} resource rather than a short-term balancing asset. This aligns with hydrogen EMS literature, which stresses that hydrogen’s value depends on coordinated operation with renewables and batteries, and on the system objectives (cost, emissions, autonomy) \cite{van2023review}.

From a policy perspective, the results support a pragmatic “fit-for-purpose hydrogen” view. Hydrogen is most defensible when long-duration flexibility is required (e.g., higher outage exposure) and/or when emissions-cost signals are strong, whereas battery-centric solutions may remain sufficient for less reliability-stressed cases. This interpretation is consistent with Australia’s strategic direction to develop a clean hydrogen industry alongside renewable electricity \cite{Australias_National_Hydrogen_Strategy_2024,Australias_National_Hydrogen_Strategy_2024_DCCEEW_Page,Australia_clean_hydrogen_opportunity_DCCEEW_2026}. It also aligns with the national emissions-reduction targets legislated through the Climate Change Act 2022, including the commitment to net zero emissions by 2050 \cite{Climate_Change_Act_2022_Parliament,Climate_Change_Act_2022_LegislationGov}.

\subsection{Barriers to deployment: infrastructure, transport, and social acceptance}
Despite strong technical potential, several barriers remain material for community-scale hydrogen-enabled microgrids.

\subsubsection{Infrastructure and logistics}
Hydrogen pathways introduce additional infrastructure (electrolyzers, storage, fuel cells, safety systems) and operations capability compared with battery-centric microgrids. Hydrogen reviews consistently highlight the importance of accurately representing real-life constraints and adopting robust EMS strategies to coordinate conversion and storage efficiently \cite{abbas2023hydrogen,ali2024fuelcell}. In remote settings, transport and maintenance logistics can become binding constraints, reinforcing the value of scenario testing prior to procurement.

\subsubsection{Regulatory readiness and social license}
Community microgrids often require local participation for governance, siting, tariff acceptance, and operational rules (especially where demand response, export limits, or community benefit-sharing are involved). Social acceptance can be a determining factor for siting and operating energy infrastructure, particularly for novel assets such as hydrogen storage. In practice, clear safety standards, transparent risk communication, and community engagement processes are needed to ensure project bankability and long-term acceptance.

\subsection{Opportunities: community-led microgrids and P2P energy trading}
The results also highlight opportunities to combine robust planning with community-led implementation models and local market mechanisms to improve both affordability and acceptance.

\subsubsection{Community-led microgrids}
Australia is actively supporting regional and remote microgrids, including First Nations contexts, to improve local energy reliability and affordability while accelerating decarbonisation \cite{ARENA_Regional_Microgrids_Program_2023,ARENA_First_Nations_microgrids_2025}. Such initiatives can de-risk early projects, build local capability, and accelerate replication---which aligns with the finding that financing conditions and reliability assumptions strongly shape optimal design outcomes.

\subsubsection{P2P trading and coordinated operation}
Peer-to-peer trading and coordinated multi-agent operation can improve utilisation of distributed assets and enable energy sharing across community participants, particularly when coupled with demand response and forecasting-aware scheduling \cite{arevalo2025systematic,raj2025review,rodriguez2024energy}. For high-renewable systems, these mechanisms can reduce curtailment, improve economic performance, and support fairness via transparent allocation of benefits.

\subsection{Alignment with Australian policy and investment pathways}
The sensitivity results indicate that microgrid economics are highly responsive to cost of capital and policy certainty. Consequently, public finance mechanisms and structured deployment programs can materially influence feasibility and replication. Australia’s hydrogen and decarbonisation policies provide a supporting context for renewable-dominant microgrids with optional hydrogen pathways \cite{Australias_National_Hydrogen_Strategy_2024,Australia_clean_hydrogen_opportunity_DCCEEW_2026,National_Hydrogen_Strategy_MediaRelease_2024}. Further, climate investment mechanisms can reduce financing barriers for capital-intensive clean energy infrastructure and improve viability of high-renewable designs \cite{CEFC_home_2026}. Finally, alignment with the Climate Change Act 2022 strengthens the rationale for sensitivity testing around emissions costs and carbon pricing; however, translating modelling results into policy decisions requires realistic boundary conditions (e.g., export constraints and emissions accounting rules) to avoid overestimating economic benefits \cite{Climate_Change_Act_2022_Parliament,Climate_Change_Act_2022_LegislationGov}.

\section{Conclusion}
This paper presented an integrated techno-economic framework for optimal microgrid design under uncertainty and demonstrated its application to a 1000-household residential community case study in Rockhampton, Queensland (Australia). The framework combines chronological simulation, lifecycle economic assessment, and structured sensitivity analysis to support decision-oriented microgrid planning. A hybrid architecture was evaluated, including PV, wind, battery storage, diesel backup and grid exchange, with an optional hydrogen pathway (electrolyzer–hydrogen storage–fuel cell) to provide long-duration flexibility.

The results highlight that the “optimal” configuration is strongly shaped by assumptions that are typically uncertain at the planning stage. Across the tested sensitivity dimensions—discount rate, technology capital costs, diesel fuel price, load demand uncertainty, renewable resource variability, carbon pricing/emissions cost, and grid outage duration—the preferred sizing and cost outcomes can shift markedly, often exhibiting regime changes where capital-intensive renewable–storage expansion becomes economically attractive under certain financial and policy conditions. Reliability stress tests further show that reduced grid availability increases the value of local generation and storage capacity, reinforcing the importance of explicitly modelling outage conditions when planning community microgrids. The no-hydrogen comparison provides attribution insight, demonstrating how excluding hydrogen alters the balance between renewable generation, battery capacity, and backup reliance, and clarifies when hydrogen’s long-duration capability provides value beyond battery-centric solutions.

Overall, the study confirms that robust microgrid planning requires sensitivity-driven design rather than relying on a single deterministic optimum. Future work will extend the framework by incorporating higher-resolution time steps, explicit network constraints and export limits, improved emissions accounting, and advanced EMS/forecasting-based dispatch strategies, enabling more realistic representation of operational constraints and policy mechanisms in Australian community microgrids.

\section*{Acknowledgment}
Authors would like to acknowledge the funding received from Australian Research Council (ARC) Linkage Project (LP190101251) titled “Advanced MGs for Residential and Commercial and Industry Buildings”


\section*{Data Availability}
Data will be made available on request.

\bibliographystyle{model1-num-names}

\bibliography{ref}

\end{document}